\documentclass[journal]{IEEEtran}
\usepackage{amsmath,amsfonts}
\usepackage{amsthm,amssymb}
\usepackage[ruled,linesnumbered,vlined]{algorithm2e}
\usepackage{array}
\usepackage{subfigure}
\usepackage{textcomp}
\usepackage{stfloats}
\usepackage{url}
\usepackage{verbatim}
\usepackage{graphicx}
\usepackage{cite}
\usepackage{xcolor}

\newtheorem{theorem}{Theorem}
\newtheorem{definition}{Definition}
\newtheorem{lemma}{Lemma}
\newtheorem{proposition}{Proposition}

\usepackage{float}
\usepackage{setspace}

\SetKwRepeat{Do}{do}{while}

\begin{document}

\title{Adaptive Privacy-Preserving Coded Computing With Hierarchical Task Partitioning}


\author{Qicheng Zeng, Zhaojun Nan, Sheng Zhou
\thanks{
Qicheng Zeng, Zhaojun Nan, and Sheng Zhou are with the
Beijing National Research Center for Information Science and Technology,
Department of Electronic Engineering, Tsinghua University, Beijing 100084,
China (e-mail: zengqc19@mails.tsinghua.edu.cn; nzj660624@mail.tsinghua.edu.cn;
sheng.zhou@tsinghua.edu.cn).
}
}



\maketitle

\begin{abstract}
Distributed computing is known as an emerging and efficient technique to support various intelligent services, such as large-scale machine learning.
However, privacy leakage and random delays from straggling servers pose significant challenges. To address these issues, coded computing, a promising solution that combines coding theory with distributed computing, recovers computation tasks with results from a subset of workers. In this paper, we propose the adaptive privacy-preserving coded computing (APCC) strategy, which can adaptively provide accurate or approximated results according to the form of computation functions, so as to suit diverse types of computation tasks. We prove that APCC achieves complete data privacy preservation and demonstrate its optimality in terms of encoding rate, defined as the ratio between the computation loads of tasks before and after encoding.
To further alleviate the straggling effect and reduce delay, we integrate hierarchical task partitioning and task cancellation into the coding design of APCC. The corresponding partitioning problems are formulated as mixed-integer nonlinear programming (MINLP) problems with the objective of minimizing task completion delay. We propose a low-complexity maximum value descent (MVD) algorithm to optimally solve these problems. Simulation results show that APCC can reduce task completion delay by a range of $20.3\%$ to $47.5\%$ when compared to other state-of-the-art benchmarks. 
\end{abstract}


\section{Introduction}
Under the vision of ``Internet of Everything", intelligence-enabled applications are essential, leading to a variety of crucial computation tasks, such as training and inference of complex machine learning models based on extensive datasets \cite{dean2012large,abadi2016tensorflow,nguyen2019machine}. However, executing these computation-intensive tasks on a single device with limited computation capability and power resources presents significant challenges. To this end, distributed computing emerges as a practical solution, where a central node, referred to as \textit{master}, manages task division, assignment, and result collection, while multiple distributed computing nodes, called \textit{workers}, process the assigned partial computation tasks in parallel \cite{sun2022coded}.

Nevertheless, while distributed computing accelerates the computation process by employing multiple workers for parallel processing, the total delay is dominated by the slowest worker, as the master must wait for all workers to complete their assigned tasks \cite{dean2013tail}. The delay of the slowest worker can exceed five times that of others, as demonstrated in the experimental results in \cite{tandon2017gradient}. Moreover, due to the randomness of delays, slow workers are difficult to identify in advance. To tackle this so-called \textit{straggling effect}, a promising approach is to adopt coded computing \cite{lee2017speeding,tandon2017gradient,li2017fundamental,yu2017polynomial,yu2020straggler,Hierarchical,reisizadeh2019coded}, which combines coding theory with distributed computing. By adding computational redundancies during the encoding process, this approach allows computation tasks to be completed with results from a subset of workers.

In coded computing, workers are tasked with processing input data and returning results, which raises privacy and security concerns. Computation tasks may involve sensitive information, such as patient medical data, customer personal information, and proprietary company data \cite{raghupathi2014big,mcafee2012big}. It is essential to maintain the data \textit{privacy}. Simultaneously, ensuring \textit{security} in computation is necessary to prevent computation results from being tampered with by Byzantine (malicious) workers. Consequently, recent research in this field has aimed to develop coded computing strategies that address not only the straggling effect but also privacy and security concerns. One such approach combines additional random data insertion with prevalent coding methods like polynomial coding \cite{LCC,yang2018secure,chang2018capacity,aliasgari2020private,kim2019private,kakar2019capacity,nodehi2018entangled,yu2020entangled,chang2019upload,d2020gasp,akbari2021secure}. This method enhances the robustness of the system against straggling workers while also improving privacy and security by obscuring the original data.

In the majority of existing studies, matrix multiplication is treated as the primary application in coded computing, and its performance has been extensively validated. However, real-world computation tasks are often more diverse than mere matrix multiplications. For instance, in a linear regression task, the iterative process of solving weights involves calculating previous weights multiplied by the quadratic power of the input data. This implies that the coded computing scheme for matrix multiplication must be executed twice in each step, and the computation becomes considerably more complex when considering other tasks, such as inference with deep neural networks.

In terms of extending the applicability of coded computing, one of the state-of-the-art approaches is Lagrange Coded Computing (LCC) \cite{LCC}. LCC employs Lagrange polynomial interpolation to transform the input data before and after encoding into interpolation points on the computation function. This allows the recovery of desired results through the reconstruction of the interpolation function, while guaranteeing the privacy of the input data. LCC is compatible with various computation tasks, ranging from matrix multiplication to polynomial functions, and offers an optimal recovery threshold concerning the degree of polynomial functions. In \cite{akbari2021secure,9174370,nodehi2018entangled}, the problem of using matrix data as inputs and polynomial functions as computation tasks is also explored.

However, LCC still suffers from several shortcomings \cite{BACC}. First, the recovery threshold is proportional to the degree of polynomial functions, which can be prohibitively large for complex tasks. Second, Lagrange polynomial interpolation can be ill-conditioned, making it challenging to ensure the numerical stability. In \cite{BACC}, Berrut's Approximated Coded Computing (BACC) is proposed to address these shortcomings and further expand the scope of computation tasks to arbitrary functions. However, BACC only yields approximated computing results and does not guarantee data privacy preservation. Other related works \cite{jahani2021codedsketch,soleymani2022approxifer,fahim2021numerically,ramamoorthy2021numerically,charalambides2020numerically} also focus on approximate results while attempting to maintain the numerical stability of coded computing.

In the aforementioned studies, numerous aspects of coded computing have been explored; however, there is still room for improvement concerning its inherent approach to addressing the straggling effect, i.e., trading computational redundancy for reduced delay. This is because the results from straggling workers are entirely disregarded, and thus the computation resources of workers may be wasted. In \cite{Hierarchical}, a hierarchical task partitioning structure is proposed, wherein divided tasks are further partitioned into multiple layers, and workers process their assigned tasks in the order of layer indices. Consequently, straggling workers can return the results of lower layers instead of none, while fast workers can reach higher layers and return more results. Similar performance improvements are achieved by multi-message communications (MMC) \cite{buyukates2020timely,hasirciouglu2021bivariate,ozfatura2020straggler}, where workers are permitted to return partial results of assigned tasks in each time slot, enabling straggling workers to contribute to the system.

Essentially, there are three ways to alleviate the straggling effect given the total number of workers. First, the recovery threshold of coded computing schemes should be minimized, as a smaller recovery threshold implies fewer workers to wait for \cite{LCC,dutta2019optimal,yu2020straggler,yang2018secure,yu2017polynomial,yang2021coded,tang2022adaptive,soleymani2021list}. As a result, the master can recover desired computing results even with more straggling workers. Second, the computation load for each worker should be carefully designed based on its computation capability, which is formulated as optimization problems in \cite{sun2022coded,zhang2021coded,wu2020latency,van2021joint,kim2020optimal}. This approach narrows the gap between the delays of fast and slow workers. Third, workers should be capable of returning partial results of assigned tasks, allowing them to complete varying amounts of computation based on their computing capabilities, rather than fast workers completing all tasks while slow workers contribute virtually nothing. The third point aligns with the focus of hierarchical task partitioning structure and MMC.

In this work, we consider a distributed system with one master and multiple workers, and propose an adaptive privacy-preserving coded computing (APCC) strategy, which primarily focuses on the applicability for diverse computation tasks, the privacy preservation of input data, and the mitigation of straggling effect. Subsequently, the hierarchical task partitioning structure is introduced into APCC, and based on this, we propose an operation called \textit{cancellation} to prevent slower workers from processing completed tasks, which reduces resource wastes and thus improves the delay performance. Specifically, the main contributions are summarized as follows:
\begin{itemize}
    \item We propose the APCC framework that effectively mitigates the straggling effect and fully preserves data privacy. APCC is suitable for diverse computation tasks including polynomial functions and non-polynomial functions, and can adaptively provide accurate results or approximated results with controllable error.
    
    \item We rigorously prove the complete privacy preservation of input data in APCC, as well as the optimality of APCC with respect to the encoding rate. The encoding rate is defined as the ratio between the computation load of tasks before and after encoding, serving as an indicator of the performance of coded computing schemes in mitigating the straggling effect.
    
    \item
    Accounting for the randomness of task completion delay, we formulate hierarchical task partitioning problems, with or without cancellation, as mixed integer nonlinear programming (MINLP) problems with the objective of minimizing task completion delay. We propose a maximum value descent (MVD) algorithm to optimally solve the problems with linear complexity.
    
    \item
    Extensive simulations demonstrate the improvements in delay performance offered by APCC when compared to other state-of-the-art coded computing benchmarks. Notably, APCC achieves a reduction in task completion delay ranging from $20.3\%$ to $47.5\%$ compared to LCC, LCC with MMC, and BACC. Simulations also explore the trade-off between task completion delay and the level of privacy preservation.
\end{itemize}

The remainder of the paper is structured as follows. Section II presents the system model. In Section III, we propose the adaptive privacy-preserving coded computing strategy, namely APCC. In Section IV, the performance of APCC is further analyzed in terms of encoding rate, privacy preservation, approximation error, numerical stability, communication costs, encoding and decoding complexity. In Section V, we proposed the MVD algorithm to address the hierarchical task partitioning optimization problem with or without cancellation. Simulation results are provided in Section VI, and conclusions are drawn in Section VII.

\section{System Model}
\begin{figure}
    \centering
    \includegraphics[scale = 0.13]{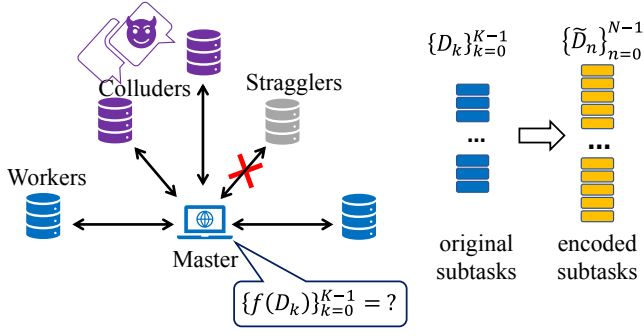}
    \caption{System Model.}
    \label{fig_Sys}
\end{figure}

In our distributed computing system, depicted in Fig. \ref{fig_Sys}, a master and $N$ workers collaborate to compute the function $f: \mathbb{V} \to \mathbb{U}$ over an equally predivided input dataset $\boldsymbol{D} = \{D_k\}_{k=0}^{K-1}$, where $D_k \in \mathbb{V}$. The master aims to obtain the results $\{f(D_k)\}_{k=0}^{K-1}$.
To achieve this, the master assigns encoded input data $\{\widetilde{D}_n\}_{n=0}^{N-1}$ to the $N$ workers for processing, awaiting their results. Subsequently, the master employs partial returned results for decoding, ultimately recovering $\{f(D_k)\}_{k=0}^{K-1}$. It is important to note that we consider the computation of $\{f(D_k)\}_{k=0}^{K-1}$ as the entire task, with the computation of $f(D_k)$ being a subtask.


Taking into account the unreliable channels and the dynamically changing computation workload of workers, some of them may fail to return results to the master in time, which are referred to as \emph{stragglers}. 
It is also assumed that workers are honest but curious, meaning they will send back the correct computation results, but there could be up to $L$ $(L<N)$ colluding workers communicating with each other and attempting to learn information about the input data $\{D_k\}_{k=0}^{K-1}$. These workers are called \emph{colluders}.




\section{Adaptive privacy-preserving coded computing}
In this section, we propose the adaptive privacy-preserving coded computing (APCC) strategy, which consists of three steps: 1) \emph{Encoding}; 2) \emph{Assignment}; 3) \emph{Decoding}. 
In the encoding and assignment steps, the $K$ equally divided input data $\{D_k\}_{k=0}^{K-1}$ are not directly encoded, instead, they are first partitioned into $r$ sets. Subsequently, the input data in each set are encoded into $N$ parts, which are then assigned to $N$ workers for computing.
In the decoding step, the master utilizes the results from the first completed subtasks in each set and interpolation methods to recover the original function $f$, achieving the purpose of decoding. 
Note that the decoding step adaptively provides accurate or approximated results according to the type of function $f$. 

In the following, we begin with a general description to explain how APCC works and then provide an illustrative example for accurate results case without loss of generality. Lastly, we introduce the hierarchical task partitioning structure of APCC, and the cancellation of completed subtasks based on this hierarchical structure.

\begin{algorithm}[t]
\caption{APCC}\label{APCC}
\KwIn{$f,D_k,r,K_i,N,L$}
\KwOut{$\{f(D_k)\}_{k=0}^{K-1}$.}
\smallskip
\textit{\textbf{1) Encoding}}:\\
The master partitions $K$ subtasks into $r$ sets, and set $i$ consists of $K_i$ subtasks $\{f(D_{i,j})\}_{j=0}^{K_i-1}$
, which satisfies $\sum_{i=0}^{r-1} K_i = K$\;

\For{$i = 0:r-1$}{
The master encodes $\{D_{i,j}\}_{j=0}^{K_i-1}$ into $\{\widetilde{D}_{i,n}\}_{n=0}^{N-1}$ according to 
$\widetilde{D}_{i,n}=g_i(\beta_n), n \in [0:N-1]$\;
}
\textit{\textbf{2) Assignment}}:\\
\For{$n = 0:N-1$}{
The master assigns $\{\widetilde{D}_{i,n}\}_{i=0}^{r-1}$ to worker $n$, and
$\{f(\widetilde{D}_{i,n})\}_{i=0}^{r-1}$ are computed in order\;
Workers return $f(\widetilde{D}_{i,n})$ to the master is once completing computation\;
}

\textit{\textbf{3) Decoding}}:\\
\For{$i = 0:r-1$}{
The master performs \textit{cancellation} for set $i$ if it is completed\;
The master decodes for set $i$ according to the first $R_i$ received results\;
\If{$f$ is  a  polynomial  function  of  degree $d$}{
Accurate results: $f(g_i(x)) = r_i(x)$\;
}
\If{$f$ is  an  arbitrary  function}{
Approximated results: $f(g_i(x)) \approx r_i(x)$\;
}
Set $i$ is completed with $f(D_{i,j})=f(g_i(\alpha_{i,j})), j\in [1:K_i]$\;
}

\end{algorithm}

\subsection{General Description}
In this subsection, we provide a general description of the proposed APCC strategy. 
As introduced in Section II, the inputs of the function $f$ are first euqally predivided into $K$ parts $D_0, D_1, \dots, D_{K-1}$, and $K$ corresponding subtasks $f(D_k), k \in [0:K-1]$ are formed. 
The APCC strategy then follows three steps: 1) \textit{Encoding}; 2) \textit{Assignment}; 3) \textit{Decoding}, and obtains accurate or approximated computing results of $\{f(D_k)\}_{k=0}^{K-1}$ according to the choice of adaptive parameter $\Tilde{w}_n$. The pseudo-code for APCC is presented in Algorithm \ref{APCC}.

\subsubsection{\textbf{\textit{Encoding}}}
In the initialization step, the $K$ subtasks are further partitioned into $r$ sets, with set $i$ containing $K_i$ subtasks ${f(D_{i,j})}_{j=0}^{K_i-1}$. Here, $K_i$ should satisfy $\sum_{i=0}^{r-1} K_i = K$. Inspired by Barycentric polynomial interpolation \cite{Barycentric,BACC}, the input data $\{D_{i,j}\}_{j=0}^{K_i-1}$ for set $i$ is linearly encoded through function $g_i(x)$ as:
\begin{align}
g_i(x) &=\sum_{j=0}^{K_i-1} \frac{\frac{w_{i,j}}{x - \alpha_{i,j}} }{\sum_{k=0}^{K_i+L-1}\frac{w_{i,k}}{x - \alpha_{i,k}}} D_{i,j} \notag \\
&+ \sum_{j=K_i}^{K_i+L-1}\frac{\frac{w_{i,j}}{x - \alpha_{i,j}} }{\sum_{k=0}^{K_i+L-1}\frac{w_{i,j}}{x - \alpha_{i,k}}} Z_{i,j},
\end{align}
where $\{Z_{i,j}\}_{j=K_i}^{K_i+L-1}$ are $L$ random matrices added to preserve the privacy, each element in $Z_{i,j}$ follows a uniform distribution, and $x \in \mathbb{R}$ is the encoding parameter.
$\{\alpha_{i,j}\}_{j=0}^{K_i+L-1}$ are distinct values selected as Chebyshev points of the first kind:
\begin{align}
\alpha_{i,j} = \cos \frac{(2j+1)\pi}{2(K_i+L)}, j\in [0:K_i+L-1].
\end{align}
$w_{i,j}$ is a constant related to $\alpha_{i,j}$ and calculated as:
\begin{align}
w_{i,j} = \frac{1}{\prod_{k=0,k \neq j}^{K_i+L-1} (\alpha_{i,j} - \alpha_{i,k})},j \in [0:K_i+L-1].
\end{align}
Note that the form of $g_i(x)$ ensures that
\begin{align}
g_i(\alpha_{i,j})=D_{i,j}, j\in [0:K_i-1].
\end{align}

The encoded input data $\{\widetilde{D}_{i,n}\}_{n=0}^{N-1}$ are obtained as:
\begin{align}
\widetilde{D}_{i,n}=g_i(\beta_n), n \in [0:N-1].
\end{align}
$\{\beta_n\}_{n=0}^{N-1}$ are selected as Chebyshev points of the second kind:
\begin{align}
\beta_n = \cos \frac{n\pi}{N-1}, n \in [0:N-1].
\end{align}

\subsubsection{\textbf{\textit{Assignment}}}
For set $i$, the encoded data $\widetilde{D}_{i,n}=g_i(\beta_n)$ is assigned to worker $n$. Consequently, each worker obtains $r$ encoded subtasks and executes them in the order of set indices. Once completed, the results of encoded subtasks $f(\widetilde{D}_{i,n})$ are returned to the master. In other words, after the original $K$ subtasks are partitioned into multiple sets, each set is transformed into $N$ encoded subtasks assigned to $N$ workers for processing. We refer to this structure as \textit{hierarchical task partitioning}.

\subsubsection{\textbf{\textit{Decoding}}}
For set $i$, the master decodes using function $r_i(x)$, which is constructed by interpolation \cite{Barycentric,BACC} as:
\begin{align}
r_i(x)=\sum_{n=0}^{R_i-1} \frac{\frac{\Tilde{w}_n}{x - \Tilde{x}_n} }{\sum_{m=0}^{R_i-1}\frac{\Tilde{w}_m}{x - \Tilde{x}_m}} f(g_i(\Tilde{x}_n)),
\end{align}
where $f(g_i(\Tilde{x}_n)), n \in [0, R_i -1]$ are the first $R_i$ received results for set $i$, 
$\Tilde{x}_n$ are the corresponding encoding parameters that belong to a subset of $\{ \beta_n, n \in [0, N-1]\}$ due to the presence of stragglers,
and the parameter $\Tilde{w}_n$ is adaptive for different cases, as follows.

\textit{\textbf{Case 1}}: \textbf{Accurate results}. If $f$ is a \textit{polynomial} function of degree $d$, where the degree $d$ of a polynomial function is defined as the maximum order of its monomials, the adaptive parameters $\Tilde{w}_n$ are determined as:
\begin{align}
\Tilde{w}_n = \frac{1}{\prod_{m=0,m \neq n}^{R_i-1} (\Tilde{x}_n - \Tilde{x}_m)},n \in [0:R_i-1].
\end{align}
In this case, $r_i(x)$ is a Barycentric polynomial interpolation function \cite{Barycentric} for $f(g_i(x))$. 
The degree of $g_i(x)$ equals $(K_i+L-1)$, so that $f(g_i(x))$ remains a polynomial function, and its degree satisfies $\mathrm{deg} f(g_i(x))\leq d (K_i+L-1)$. Consequently, if the number of received results $R_i$ for set $i$ satisfies:
\begin{align}\label{Eta_K_relation}
R_i = d (K_i+L-1)+1,
\end{align}
it implies that sufficient interpolation points have been obtained to precisely recover $f(g_i(x))$ through $r_i(x)$, and the entire computation process is completed with 
\begin{align}
    f(D_{i,j})=f(g_i(\alpha_{i,j}))=r_i(\alpha_{i,j}),
\end{align}
for any $i\in[0:r-1],j\in [0:K_i-1]$.

Note that Eq.(\ref{Eta_K_relation}) means that the accurate result case of APCC has the same recovery threshold as LCC \cite{LCC}. Besides, similar to LCC, when there is no need for privacy preservation which means $L = 0$, we can also provide an uncoded version of APCC by selecting the value of $\{\beta_n\}$ from $\{\alpha_{i,j}\}$. Thus, the new recovery threshold becomes:
\begin{align}
R_i = N - \lfloor N /K_i \rfloor +1.
\end{align}


\textit{\textbf{Case 2}}: \textbf{Approximate results}. If $f$ is an \textit{arbitrary} function, the adaptive parameter $\Tilde{w}_n$ is calculated as:
\begin{align}
\Tilde{w}_n = (-1)^n, n \in [0:R_i-1].
\end{align}
In this case, $r_i(x)$ is a Berrut's rational interpolation function for $f(g_i(x))$, as discussed in \cite{BACC,berrut1988rational}.
The computed results $f(g_i(\Tilde{x}_n))$ serve as the interpolation points of $f(g_i(x))$, and they satisfy $r_i(\Tilde{x}_n)=f(g_i(\Tilde{x}_n))$ due to the property of Berrut's rational interpolation \cite{berrut1988rational}.
Therefore, the master can regard $r_i(x)$ as an approximation of $f(g_i(x))$, which means that 
\begin{align}
f(D_{i,j}) =f(g_i(\alpha_{i,j}) \approx r_i(\alpha_{i,j}),
\end{align}
for any $i\in[0:r-1],j\in [0:K_i-1]$.
In addition, the approximation using $r_i(x)$ becomes more accurate as $R_i$ increases. Thus, if the master desires more accurate computations, it simply needs to wait for more results.




\begin{figure}
    \centering
    \subfigure[Encoding]{
    \includegraphics[scale = 0.11]{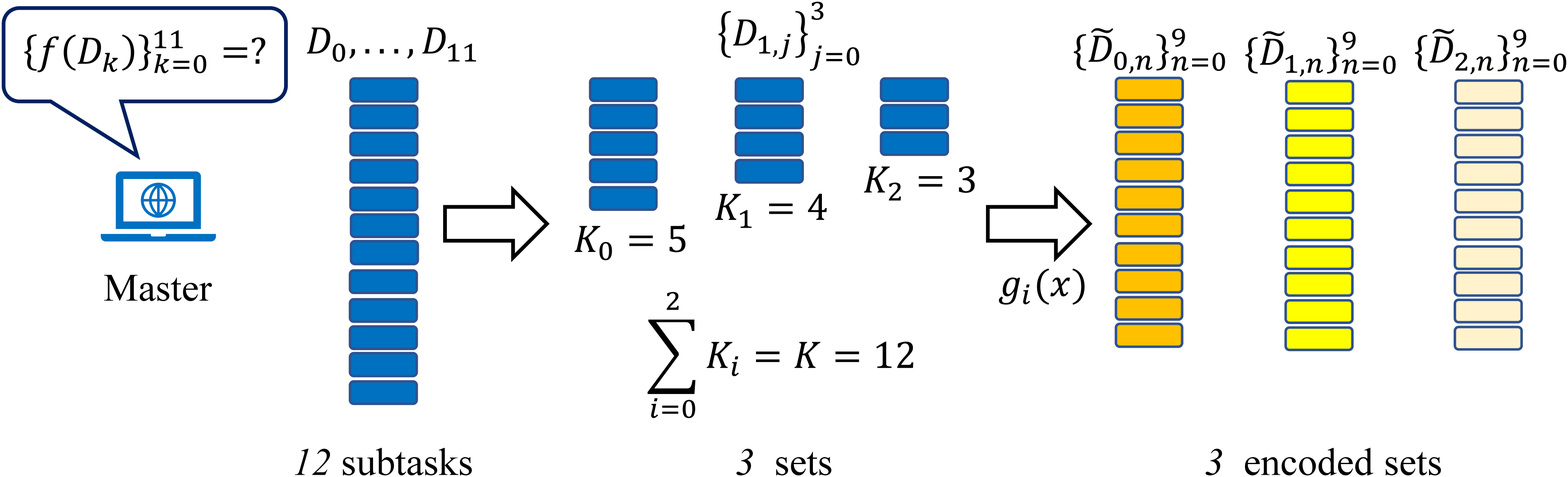}
    }
    \subfigure[Assignment and Decoding]{
    \includegraphics[scale = 0.11]{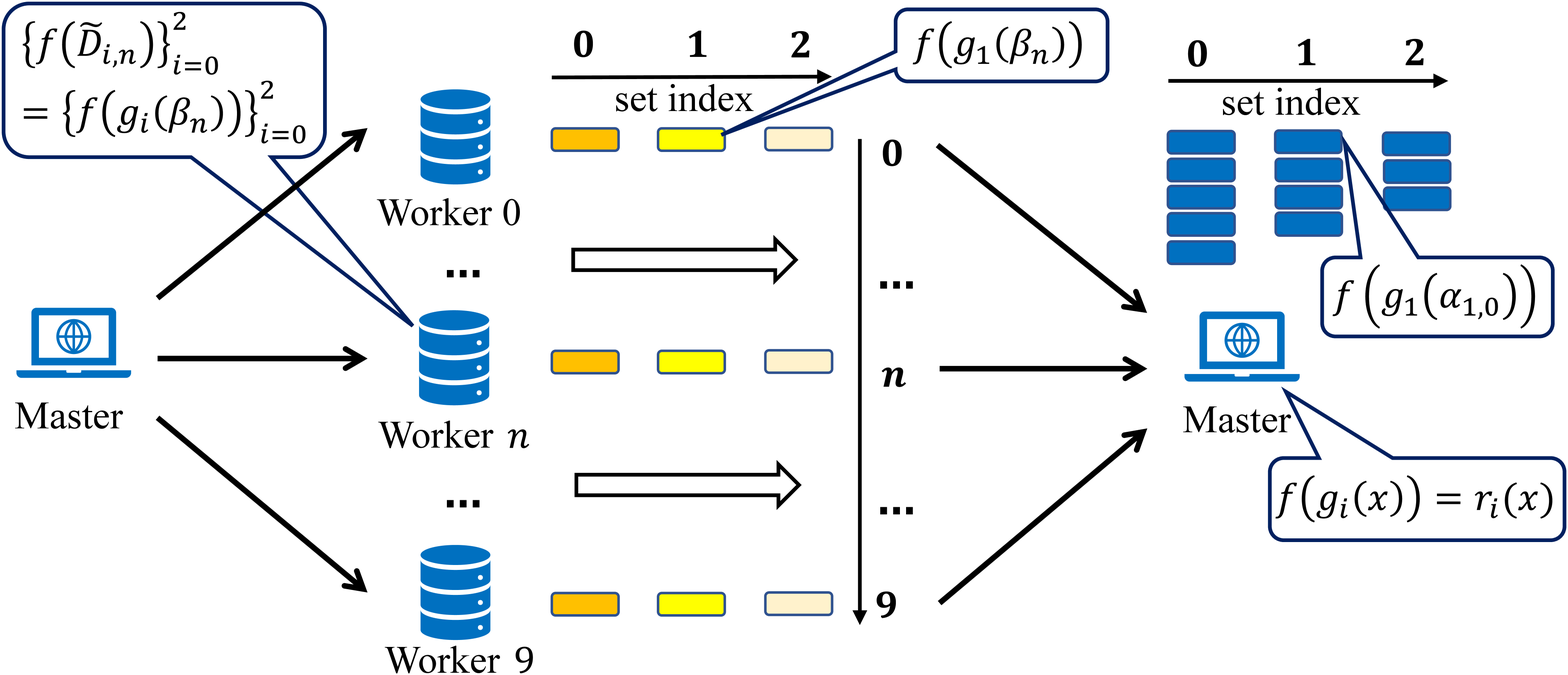}
    }
    \caption{The three-step process of APCC.}
    \label{APCC_process}
\end{figure}

\subsection{An Illustrating Example}
In this subsection, to show how APCC works, we present an illustrative example for the case of accurate results without loss of generality. Specifically, we consider a linear regression problem.
The feature data $\boldsymbol{D} \in \mathbb{R}^{p \times q}$ contains $p$ data samples with $q$ features, and the label vector is denoted by $\boldsymbol{y} \in \mathbb{R}^{p \times 1}$. The objective is to find the weighting vector $\boldsymbol{w} \in \mathbb{R}^{q \times 1}$ that minimizes the loss $||\boldsymbol{Dw-y}||^2$. To solve this problem, the gradient descent method updates the weights iteratively along the negative gradient direction as follows:
\begin{align}
\boldsymbol{w}^{(t+1)} = \boldsymbol{w}^{(t)} - \frac{2\eta}{p} \boldsymbol{D}^T (\boldsymbol{D}\boldsymbol{w}^{(t)} - \boldsymbol{y}),
\end{align}
where $\eta$ is the learning rate and $t$ represents the iteration index.

In order to apply the aforementioned update process to a distributed system with one master and $N = 10$ workers, for instance, the feature data $\boldsymbol{D}$ is first equally divided into $K = 12$ sub-matrices $(D_0, D_1, \dots, D_{11})^T$. As $\boldsymbol{w}^{(t)}$ is known by the workers and $\boldsymbol{D}^T \boldsymbol{y}$ can be precomputed by the master, the computation function (subtask) of the master in each iteration can be expressed as $f(D_k) = D_k^T D_k \boldsymbol{w}, k \in [0:11]$. After obtaining the results of the entire task $\{f(D_k)\}_{k=0}^{11}$, the gradient update is computed as $\boldsymbol{D}^T \boldsymbol{D}\boldsymbol{w} = \sum_{k=0}^{11} D_k^T D_k \boldsymbol{w}, D_k \in \mathbb{R}^{\frac{p}{12} \times q}$.



We now illustrate how APCC can be applied in the above problem, to obtain $f(D_k)=D_k^T D_k \boldsymbol{w}, k \in [0:11]$.

\subsubsection{\textbf{\textit{Encoding}}}
As depicted in Fig. \ref{APCC_process}(a), since there are 12 subtasks $f(D_k), k \in [0:11]$, the master further partitions them into $r = 3$ sets before encoding the inputs, and set $i$ ($i = 0, 1, 2$) contains $K_i$ subtasks. The value selection for ${K_i}$ will be formulated as an optimization problem in Section V. For instance, we assume that $K_0 = 5$, $K_1 = 4$, and $K_2 = 3$, and they satisfy $K_0 + K_1 + K_2 = K = 12$. After this hierarchical task partitioning, the input of the $j$-th subtask in set $i$ is denoted as $D_{i,j}$ instead of the previous $D_k$, and $\{f(D_{i,j})=D_{i,j}^T D_{i,j} \boldsymbol{w}\}_{j=0}^{K_i-1}$ are the $K_i$ subtasks in this set.

Next, the input data $\{D_{i,j}\}_{j=0}^{K_i-1}$ in set $i$ are encoded into $N = 10$ parts $\{\widetilde{D}_{i,n}\}_{n=0}^{9}$ through $g_i(x)$, where $x$ represents the encoding parameters and $\widetilde{D}_{i,n} = g_i(\beta_n), n \in [0:9]$. Moreover, $g_i(x)$ is a polynomial function with a degree of $(K_i + L - 1)$, and its form ensures that the parameters $\{\alpha_{i,j}\}$ satisfy $g_i(\alpha_{i,j})=D_{i,j}$.

\subsubsection{\textbf{\textit{Assignment}}}
As Fig. \ref{APCC_process}(b) shows, for each set, the 10 encoded input data $\{\widetilde{D}_{i,n}\}_{n=0}^9$ are assigned to the 10 workers. Subsequently, each worker applies function $f$ to compute and return the results to the master. As can be observed, the $K_i$ original subtasks in set $i$ are transformed into 10 subtasks performed on the 10 workers in parallel. Since there are 3 sets, each worker is assigned 3 subtasks. These subtasks are executed according to the order of sets, which implies $f(\widetilde{D}_{0,n})$ is computed first, followed by $f(\widetilde{D}_{1,n})$, and so on.



\subsubsection{\textbf{\textit{Decoding}}}
As illustrated in Fig. \ref{APCC_process}(b), following the assignment of encoded input to workers, the master continuously awaits the subtask results from workers, and creates a decoding function $r_i(x)$ for set $i$. This decoding function is constructed using interpolation to recover the original function $f(D_{i,j}) = f(g_i(\alpha_{i,j}))$. Consequently, each received result, $f(g_i(\beta_n))$, can be regarded as an interpolation point for $f(g_i(x))$, and $r_i(x)$ is precisely the interpolation function of $f(g_i(x))$.

Presently, $f(D_{i,j})=D_{i,j}^T D_{i,j} \boldsymbol{w}$ is a polynomial function of degree $d = 2$, where the degree $d$ of a polynomial function $f$ is defined as the maximum order of its monomials.
We have illustrated how to complete the decoding process in Subsection III.A. By setting the number of received results to $R_i = d(K_i+L-1)+1$, sufficient interpolation points are obtained to accurately recover $f(g_i(x))$ through $r_i(x)$, i.e., $f(D_{i,j})=f(g_i(\alpha_{i,j}))=r_i(\alpha_{i,j})$, for all $i\in[0:2]$ and $j\in [0:K_i-1]$.

\subsection{Hierarchical Task Partitioning and Cancellation}
In Fig. \ref{APCC_process}, the hierarchical task partitioning in APCC aims to maximize the utility of computing results from straggling workers. This is achieved through a well-designed structure and appropriate choice of $K_i$ values. Although the same number of encoded subtasks are assigned to all workers, the number of successfully returned results from each worker can differ due to varying processing speeds. As a result, straggling workers may return fewer computing results than faster workers, but they can still make valuable contributions to task completion instead of being disregarded.

Furthermore, the illustration in Fig. \ref{APCC_process} suggests that $K_{i-1}$ should exceed $K_i$ \cite{Hierarchical}. This assertion is explained as follows: The ``completion time" of set $i$" is defined as the moment when a sufficient number of encoded subtask results within set $i$ are obtained. The overarching objective is to minimize the delay in completing the entire task, which must necessarily exceed the ``completion time" of any set since the entire task remains incomplete until all $r$ sets are recovered. Given that subtasks are executed in order of set indices, when set $r$ is recovered, the master must have acquired results for the smaller-index sets equal to or greater than $K_i$. Opting for smaller values of $K_i$ for the smaller-index sets would result in more workers experiencing straggling, a situation that should be averted. Further details are expounded in Section V.

\begin{figure}
    \centering
    \includegraphics[scale = 0.12]{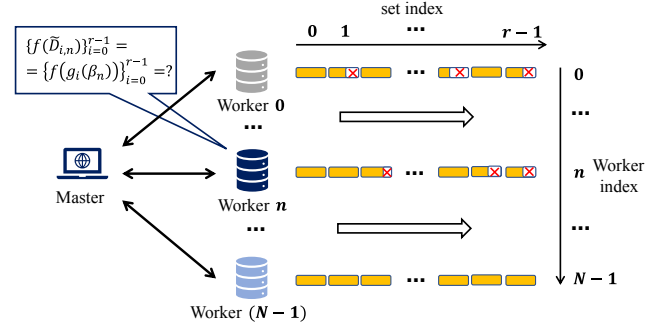}
    \caption{Hierarchical structure and the cancellation operation.}
    \label{HierarchicalAndCancellation}
\end{figure}

Based on the hierarchical structure, we propose an alternative method to further accelerate the coded computing process. As depicted in Fig. \ref{HierarchicalAndCancellation}, the subtasks to be computed on each worker form an execution sequence. Once enough results for set $i$ are obtained, the master can instruct workers that have not completed the computation of $f(\widetilde{D}_{i,n})$ to terminate or skip this part of the computation and proceed to compute the next subtasks $f(\widetilde{D}_{i+1,n})$ of the subsequent set. This operation, called        ``\textit{Cancellation}", prevents computation resources from being wasted on completed sets. Considering the presence of non-persistent stragglers, cancellation increases the probability of them overcoming the previous straggling effect and avoiding becoming stragglers again. 

\section{Performance analysis}
In this section, we first define a metric called \textit{encoding rate} to evaluate the efficiency performance of coded computing schemes, in terms of utilizing computation resources of workers as efficiently as possible. Then based on the optimal recovery threshold of LCC \cite{LCC}, we rigorously prove APCC for accurate results is also an optimal polynomial coding in terms of the encoding rate. Furthermore, an information-theoretic guarantee to completely preserve the privacy of input data $\{D_k\}_{k=0}^{K-1}$ in APCC is proved. Subsequently, we present an analysis of the approximation error for Case 2 of APCC, along with a discussion of numerical stability. At the end of this section, we provide a detailed analysis about the communication costs, encoding and decoding complexity for APCC, and compare it with other state-of-the-art strategies.

\subsection{Optimality of APCC in Terms of Encoding Rate}

To evaluate the performance of various coded computing schemes, a metric known as the encoding rate $R_{\rm encode}$ is used. This metric is defined as:
\begin{align} \label{R_encode}
R_{\rm encode} = \frac{K}{N-S},
\end{align}
where $K$ is the number of subtasks before encoding, $N$ is the number of subtasks after encoding (which is equivalent to the number of workers), and $S$ represents the number of straggling workers that failed to return results before the task was completed. Similar metrics, such as those found in \cite{chang2018capacity, kakar2019capacity, zeng2021capacity}, have also been developed. 

Besides, since the recovery threshold, denoted by $H$, is defined as the minimum number of results needed to guarantee decodability, we have $H = N-S$ and thus $R_{\rm encode} = K/H$.
It is important to note that the encoding rate only applies when decodability is guaranteed.

The physical significance of the encoding rate is the ratio between the computation load of tasks before encoding and that required after encoding. For instance, given a task with a computation load of $O(\gamma)$, each subtask has a corresponding load of $O(\frac{\gamma}{K})$. As $(N-S)$ subtasks are successfully completed, the required computation load is $O(\frac{\gamma(N-S)}{K})$.
Since coded computing essentially trades computation redundancy for reduced delay to mitigate the straggling effect, it is reasonable to use this metric to evaluate the efficiency of different schemes.

Before proving the optimality of APCC, we present the definitions of capacity and linear coded computing schemes.
\begin{definition}
A linear coded computing scheme is one in which the encoded data is a linear combination of the original input data as follows:
\begin{align}
\widetilde{D}_n = \sum_{k=0}^{K-1} G_{n,k} D_k +\widetilde{Z}_n, n\in[0:N-1],
\end{align} 
where $G \in \mathbb{R}^{N \times k}$ is the encoding matrix corresponding to the aforementioned encoding function $g_i(x)$, and $\widetilde{Z}_n$ is the additive randomness.
\end{definition}
\begin{definition}
The capacity $C$ for a distributed computing system with one master and $N$ workers,  where the computation function used by the master is $f$, is defined as the supremum of the encoding rate $R_{\rm encode}$ as:
\begin{align}
C \triangleq \sup R_{\rm encode},
\end{align}
over all feasible linear coded computing schemes that can address up to $L$ colluders and $S$ stragglers
\end{definition}

As illustrated in Section III, APCC is a linear coded computing scheme and its hierarchical structure results in different $K_i$ and $S_i$ for each set, with $K_i$ and $S_i$ representing the number of subtasks before encoding and that of straggling workers, respectively. For set $i$, $R_i$ represents the number of workers that have successfully returned results in time, implying that the number of stragglers is $S_i=N-R_i$. Moreover, set $i$ is considered complete when $R_i = d (K_i+L-1)+1$.
Hence, the encoding rate of APCC can be calculated as:
\begin{align}\label{R_encode_1}
R_{\rm encode}^{[\rm APCC]} = \frac{K_i}{N-S_i} = \frac{N-S_i-d(L-1)-1}{d(N-S_i)},
\end{align}
or the uncoded version for $L = 0$ :
\begin{align}\label{R_encode_2}
R_{\rm encode}^{[\rm APCC]} = \frac{K_i}{N-S_i} \leq \frac{N}{(N-S_i)(S_i+1)},
\end{align}
where the equality holds when $N$ can be divided by $K_i$.

The following theorem shows that the encoding rate of APCC achieves the capacity, thereby establishing the optimality of APCC. For the sake of clarity, we omit the set index $i$ and focus on a specific set, without loss of generality.
\begin{theorem}
For a coded computing problem $(N,S,L,f)$, where $N$ is the number of workers, $S$ and $L$ denote the number of stragglers and colluders, respectively, and $f$ is an arbitrary polynomial function of degree $d$, the capacity $C$ is given by:
\begin{align}\label{Capacity_theorem}
    C = \left\{
    \begin{array}{ll}
         \frac{N-S-d(L-1)-1}{d(N-S)}, & {\rm if} \,\, L>0,  \\
         \max\{\frac{N-S+d-1}{d(N-S)}, \, \frac{N}{(N-S)(S+1)}\} & {\rm if} \,\, L=0.
    \end{array}  \right.
\end{align} 
\end{theorem}

Essentially, the optimality of APCC in encoding rate is attributed to its identical polynomial coding structure when compared to LCC \cite{LCC}, despite having distinct function expressions. This observation also implies that the encoding rate of LCC can similarly achieve the capacity.

To prove Theorem 1, 
a lower bound on the capacity $C$ is first established,
which follows the encoding rate of APCC in (\ref{R_encode_1}) and (\ref{R_encode_2}). To establish the upper bound, we leverage the optimality statement of LCC, as illustrated in Theorem 1 and 2 of \cite{LCC}, which shows that the decodability of polynomial coded computing is guaranteed only when the following condition is met:
\begin{align} \label{LCC_optimality}
    N \geq \left\{
    \begin{array}{ll}
        d(K+L-1)+1+S, & {\rm if} \,\, L>0,  \\
         \min\{d(K-1)+1+S, \, K(S+1)\} & {\rm if} \,\,   L=0.
    \end{array}  \right.
\end{align} 
Therefore, we have:
\begin{align} \label{K Constr}
    K \leq \left\{
    \begin{array}{ll}
         \frac{N-S-1}{d} - L+1, & {\rm if} \,\, L>0,  \\
         \max\{\frac{N-S+d-1}{d}, \, \frac{N}{(S+1)}\} & {\rm if} \,\, L=0,
    \end{array}  \right.
\end{align} 

Ineq.(\ref{K Constr}) presents the maximum number of tasks divisions permissible  to ensure decodability, given the numbers of workers $N$, stragglers $S$ and colluders $L$.
The reason is that the more divisions there are, the more results are needed from workers. However, there are at most $N$ workers, including $S$ stragglers, to return results. Based on (\ref{K Constr}), an upper bound on the encoding rate can be derived as:
\begin{align}
    &R_{\rm encode} = \frac{K}{N-S} \notag \\
    &\leq \left\{
    \begin{array}{ll}
         \frac{N-S-d(L-1)-1}{d(N-S)}, & {\rm if} \,\, L>0,  \\
         \max\{\frac{N-S+d-1}{d(N-S)}, \, \frac{N}{(N-S)(S+1)}\} & {\rm if} \,\, L=0.
    \end{array}  \right.
\end{align} 
Since the capacity $C$ is the supremum of $R_{\rm encode}$, it also has the same upper bound. With the lower bound provided previously, we can conclude that APCC is an optimal coded computing strategy that can reach the capacity in (\ref{Capacity_theorem}).


To enhance clarity, the fundamental proof for the derivation of (\ref{LCC_optimality}) is briefly introduced in Appendix A, following the same steps as outlined in \cite{LCC}. Nevertheless, we present an alternative and more concise proof for the construction of the involved multilinear function.




Please note that the conclusion presented in this subsection pertains only to accurate coded computing. For approximated coded computing, the use of different approximation methods can lead to varying errors, making it challenging to compare and analyze their impact on the encoding rate and capacity in a qualitative manner.


\subsection{Guarantee of the Privacy Preservation}
Recall that colluders are those workers who can communicate with each other and attempt to learn something about the original input data. Since the system can tolerate at most $L$ colluders, we assume that there are $L'$ colluders, where $L' \leq L$, and the user does not know which workers are colluding. We use the index set $\mathcal{L}= \{ l_0, l_1, \dots, l_{L'-1} \} \subseteq \{0,\dots,N-1\}$ to denote the colluding workers, where $|\mathcal{L}|=L'$.

Assuming that the input data $\{D_{i,j}\}_{j=0}^{K_i-1}$ are independent of each other, we denote the encoded input data sent to workers in the colluding set $\mathcal{L}$ for set $i$ as:
\begin{align}
\widetilde{D}_{i,\mathcal{L}} \triangleq (\widetilde{D}_{i,l_0}, \widetilde{D}_{i,l_1}, \dots, \widetilde{D}_{i,l_{L'-1}} ).
\end{align}
Therefore, the information-theoretic privacy-preserving constraint can be expressed as:
\begin{equation}
I(D_{i,0}, D_{i,1}, \dots, D_{i,K_i-1};\widetilde{D}_{i,\mathcal{L}}) =0, \forall i\in[0,r-1],
\end{equation}
where $I(\cdot)$ represents the mutual information function.

With the assumption of finite precision floating point arithmetic, the values of elements in the data matrices such as $D_{i,j}$, $\widetilde{D}_{i,n}$, and $Z_{i,j}$ come from a sufficiently large field $\mathbb{F}$. Given that the size of these data matrices is $m\times m'$, we have
\begin{align}
    & I(D_{i,0}, D_{i,1}, \dots, D_{i,K_i-1};\widetilde{D}_{i,\mathcal{L}}) =H(\widetilde{D}_{i,l_0}, \dots, \widetilde{D}_{i,l_{L'-1}}) \notag\\
    &- H(\widetilde{D}_{i,l_0}, \dots, \widetilde{D}_{i,l_{L'-1}}|D_{i,0},\dots,D_{i,K_i-1}) \notag \\
    &\stackrel{(a)}{=} H(\widetilde{D}_{i,l_0}, \dots, \widetilde{D}_{i,l_{L'-1}}) - H(Z_{i,K_i},\dots,Z_{i,K_i+L-1}) \notag \\
    &\stackrel{(b)}{=} H(\widetilde{D}_{i,l_0}, \dots, \widetilde{D}_{i,l_{L'-1}}) - L  m m'  \log |\mathbb{F}| \notag \\
    &\leq H(\widetilde{D}_{i,l_0})+ \cdots + H(\widetilde{D}_{i,l_{L'-1}})- L  m m' \log |\mathbb{F}|\notag \\
    &\stackrel{(c)}{\leq} L'  m m'  \log |\mathbb{F}|-  L  m m'  \log |\mathbb{F}|
     = 0, \forall i\in[0,r-1], \label{Mutual Info Constr}
\end{align}
where $(a)$ is due to the fact that all random matrices $\{Z_{i,j}\}_{j=K_i}^{K_i+L-1}$ are independent of the input data $\{D_{i,j}\}_{j=0}^{K_i-1}$. $(b)$ is because the entropy of each element in the random matrices equals $\log |\mathbb{F}|$, and $(c)$ follows from the upper bound of the entropy of each element in $\widetilde{D}_{i,l_{(\cdot)}}$ being $\log |\mathbb{F}|$. Since the mutual information is non-negative, it must be 0, which guarantees complete privacy preservation.

Note that the analysis in this subsection is applicable to both accurate and approximate cases. This is because that the analysis only involves the encoding and assignment steps of APCC, and both cases require the same two initial steps. The key difference between the two aforementioned cases is reflected in the decoding functions with distinct adaptive parameters $\Tilde{w}_n$, which correspond to Barycentric polynomial interpolation and Berrut's rational interpolation, respectively.

\subsection{Analysis of Approximation Error for Case 2}
According to the discussion in \cite{BACC}, let the interpolating objective function $h_i(x)=f(g_i(x))$ has a continuous second derivative on $[-1,1]$, and the number of received results $R_i > 3$, the approximation error is upper bounded as:
\begin{align}
    &||r_i(x)-h_i(x)|| \leq \notag\\
     &2 (1+\Gamma) \sin \frac{(N-R_i+1)\pi}{2(N-1)} ||h_i''(x)||,
\end{align}
if $R_i$ is even, and
\begin{align}
    &||r_i(x)-h_i(x)|| \leq \notag\\
    &2 (1+\Gamma) \sin \frac{(N-R_i+1)\pi}{2(N-1)} (||h_i''(x)||+||h_i'(x)||),
\end{align}
if $R_i$ is odd, where $\Gamma \triangleq \frac{(N-R_i+1)(N-R_i+3)\pi^2}{4}$.

Consequently, for set $i$ and a fixed total number of workers $N$, the approximation using $r_i(x)$ becomes more accurate as the number of received results $R_i$ increases.

\subsection{Numerical Stability}
In coded computing, the issue of numerical stability typically arises from the decoding part, which is based on solving a system of linear equations involving a Vandermonde matrix. As previously discussed, Cases 1 and 2 of APCC employ Barycentric polynomial interpolation and Berrut's rational interpolation as decoding methods, respectively.
For Case 1, Barycentric polynomial interpolation demonstrates good performance in addressing errors caused by floating-point arithmetic \cite{Barycentric}. Regarding Case 2, it has been shown in \cite{BACC} that the Lebesgue Constant of Berrut's rational interpolation grows logarithmically with the number of received results from workers, rendering it both forward and backward stable.

\subsection{Communication Costs}
In this subsection, we analyze the communication costs of APCC concerning the number of partitioning sets $r$ and the number of task divisions $K$. For comparison, we also provide the analysis of communication costs for LCC \cite{LCC}.
The communication costs refers to the bit size of transmission data including encoded input data and required computing results. Given the computation function $f$ and the original input data $\{D_k\}_{k=0}^{K-1}$, we assume that the communication costs to transmit $\{D_k\}_{k=0}^{K-1}$ are $O(x)$ for all strategies, and those to transmit $\{f(D_k)\}_{k=0}^{K-1}$ back to the master are $O(\sigma x)$.
\subsubsection{APCC}
Since the entire task is divided into $K$ subtasks in APCC, and each worker (totally $N$ workers) is assigned $r$ subtasks, the costs for each worker are $O\left(\frac{xr}{K}\right)$. Therefore, the total communication costs of encoded input data are $O\left(\frac{xrN}{K}\right)$. On the other hand, the master needs to receive $\sum_{i=0}^{r-1} \left[d (K_i + L -1 ) + 1 \right]= d(K +rL-r)+r$ results to complete the entire task for Case 1 of APCC, so the total communication costs of feedback is $O\left(\frac{\sigma x}{K} \left[ d(K +rL-r)+r \right]\right)$. For Case 2 of APCC, the costs per feedback result are $O\left(\frac{\sigma x}{K}\right)$. Additionally, when considering cancellation, the master only needs to send signals to all workers at the moment when one set is finished, so the signaling overhead is relatively small compared to the communication costs of the computation data.
\subsubsection{LCC}
According to \cite{LCC}, LCC divides the entire task into $K'$ subtasks $\{f(D_k)\}_{k=0}^{K'-1}$, and each worker is only assigned with one encoded subtask, so the costs for each worker in LCC are $O(\frac{x}{K'})$. Consequently, the total communication costs of encoded input data are $O(\frac{xN}{K'})$. In LCC, the master needs to receive $d(K' + L -1)+1$ results to complete the entire task. Therefore, the total communication costs of feedback for LCC is $O\left(\frac{\sigma x}{K'} \left[ d(K' +L-1)+1 \right]\right)$. 



Note that when comparing the delay performance of different coded computing strategies, we should make the computation loads for a single worker the same to ensure fairness. Assuming the computation loads of subtasks are proportional to the bit size of input data, which are exactly the communication costs for each worker, we have
$O\left(\frac{x}{K'}\right) = O\left(\frac{xr}{K}\right)$,
and thus we can derive
\begin{align}
K' = K/r.
\end{align}
Accordingly, the communication costs of both encoded input data and feedback for the two strategies are the same. This is reasonable because, despite APCC requiring the reception of more feedback results due to multiple sets, the bit size of an individual result is smaller compared to LCC, thanks to a more granular task division where $K=K'r$.

In fact, the reduction of computation delay through the hierarchical task partitioning structure in APCC does not entail increased communication costs. Rather, it is primarily attributed to its capability to enable workers to return partial results as they return every result of subtasks, thereby achieving more efficient utilization of the computing resources of the straggling workers. However, the partitioning of multiple sets in APCC does lead to increased encoding and decoding times, as elucidated in the subsequent subsection.

\subsection{Encoding and Decoding Complexity}
In this subsection, we provide the analysis of encoding and decoding complexity.
Intuitively, APCC utilizes the hierarchical task partitioning structure to enhance delay performance. However, it does so at the cost of requiring multiple encoding and decoding operations, specifically $r$ times for the $r$ sets, when compared to LCC \cite{LCC} and BACC\cite{BACC}. 

In LCC and BACC, the encoding operations take $N$ times, corresponding to the number of workers, while the decoding operations take $K'$ times, equivalent to the number of task divisions.
On the other hand, in the case of APCC, which features $r$ partitioned sets, the encoding and decoding operations entail $Nr$ and $\sum_{i=0}^r K_i = K$, respectively. When the computation loads per worker in all strategies are equal, i.e., $K' = K/r$, it can be deduced that the encoding and decoding operations in APCC are $r$ times those of LCC and BACC.


\section{Hierarchical Task partitioning}
In this section, the hierarchical task partitioning is formulated as an optimization problem with the objective of minimizing the task completion delay. The problem is divided into two cases for consideration: with and without cancellation. Through derivations, two mixed integer programming problems are obtained, and we propose a maximum value descent (MVD) algorithm to obtain the optimal solutions with low computational complexity. Moreover, after analysis, it is found that the MVD algorithm can be quickly executed by selecting appropriate input. Detailed explanations are provided as follows.

\subsection{Problem Formulation}
In the context of negligible encoding and decoding delays, with the computation delays of workers being the dominant component, the delay for a worker to complete a single subtask, denoted as $T$ can be represented by a shifted exponential distribution
\cite{lee2017speeding,reisizadeh2019coded,Hierarchical,zhang2021coded,wu2020latency,sun2022coded}, whose cumulative distribution function (CDF) is given by:
\begin{align}\label{DelayT}
    F_T(t) = \mathbb{P}[T \leq t] = \left\{
    \begin{array}{ll}
         1 - e^{-\mu (t - a)}, & {\rm if} \,\, t \geq a,  \\
         0 & {\rm otherwise},
    \end{array}  \right.
\end{align} 
where $a > 0$ is a parameter indicating the minimum processing time and $\mu > 0$ is a parameter modeling the computing performance of workers. All $N$ workers follow a uniform computation delay distribution defined in (\ref{DelayT}).

Recall that in the hierarchical structure, the completion of a particular set is dependent on the successful receiving of a sufficient number of results from its encoded subtasks. The overall completion of the entire task is achieved only when all $r$ sets have been completed. Notably, $H_i$ is defined as the threshold number of successful results needed to ensure the completion of set $i$.

Following the discussion in Section III and assuming that privacy preservation is required which means $L>0$, the threshold for \textit{Case} 1 of APCC can be expressed as $H_i = d(K_i+L-1)+1$ according to (\ref{Eta_K_relation}). For \textit{Case} 2 of APCC, the threshold $H_i$ can be determined based on the desired approximation precision, with higher values of $H_i$ leading to more accurate approximations.


The completion time of sets is defined as $\boldsymbol{t} \triangleq \{t_i, i \in [0:r-1]\}$, where $t_i$ denotes the time interval from the initial moment 0 of the entire task to the recovery moment of set $i$. The entire task is considered completed when all $r$ sets have been recovered. Therefore, we denote the entire task completion delay as
\begin{equation} 
T^{[e]} = \max_{i\in[0:r-1]} t_i. 
\end{equation}
Note that while each worker executes the assigned subtasks in the order of set indices, the order in which these sets are recovered may not be the same. The completion time of sets is influenced not only by the set indices but also by the recovery thresholds $H_i$ determined by $K_i$. 

Due to the randomness of delay, our objective is to minimize the entire task completion delay $T^{[e]} = \mathop{\max}_{i\in[0:r-1]} t_i$, upon which the probability of the master recovering desired results for all sets is higher than a given threshold $\rho_s$, as expressed by the following inequality:
\begin{align}
\label{Threshold_rho}
\mathbb{P}[R_0(t_0) \geq H_0, \dots, R_{r-1}(t_{r-1}) \geq H_{r-1}] \geq \rho_s,
\end{align}
where $R_i(t)$ is defined as the number of returned results for set $i$ until time $t$.

However, to derive (\ref{Threshold_rho}), we first need to obtain the distribution of the delay required to receive the last non-straggling result in each set and then derive their joint probability distribution, which is intractable, especially when considering the cancellation of completed sets. As a result, the problem with the constraint (\ref{Threshold_rho}) is hard to solve.

In the following, we consider substituting (\ref{Threshold_rho}) with an expectation constraint (\ref{inequation2}) and formulate the problem as:
\begin{subequations}
\begin{align}
    \mathcal{P}1-1:  \mathop{\min}_{\{\boldsymbol{K}\}} \, &\mathop{\max}_{i\in[0:r-1]} \, t_i\\
    \label{equation1} \mathbf{s.t.} \, & \sum_{i=0}^{r-1} K_i = K,\\
    \label{inequation1} & H_i \leq N, \forall i \in [0,r-1]\\
    \label{inequation2} & \mathbb{E}[R_i(t_i)] \geq H_i, \forall i \in [0,r-1]\\
    \label{integer1} &  K_i, H_i \in \mathbb{Z}^+, \forall i \in [0,r-1],
\end{align}
\end{subequations}
where $\boldsymbol{K} \triangleq \{K_i| i \in [0:r-1]\}$ is the partitioning scheme.

Constraint (\ref{equation1}) corresponds to the hierarchical task partitioning, and (\ref{inequation1}) indicates that the threshold for each set should be smaller than the number of workers. In constraint (\ref{integer1}), $\mathbb{Z}^+$ represents the set of positive integers. Constraint (\ref{inequation2}) states that the master is expected to receive sufficient results of encoded subtasks from workers to recover ${f(D_{i,j})}_{j=0}^{K_i-1}$ in set $i$. Similar approximation approaches are also used in \cite{reisizadeh2019coded,zhang2021coded,wu2020latency,sun2022coded}, and the performance gap can be bounded \cite{reisizadeh2019coded}.

As previously shown, $H_i = d(K_i+L-1)+1$ for \textit{Case} 1 of APCC. Additionally, the maximum of $t_i$ for all sets can be replaced with an optimization variable $z$ by adding an extra constraint. Consequently, for Case 1 of APCC, $\mathcal{P}1-1$ can be equivalently written as:
\begin{subequations}
\begin{align}
    \mathcal{P}1-2:  \mathop{\min}_{\{\boldsymbol{K}, z\}} \, z \quad \quad &\\
    \label{inequation4} \mathbf{s.t.} \quad
    t_i -z &\leq 0, \forall i \in [0,r-1],\\
    \label{inequation3} d(K_i+L-1)+1 - \mathbb{E}[R_i(t_i)] &\leq 0, \forall i \in [0,r-1],\\
    \label{inequation6} d (K_i +L -1) +1 - N &\leq 0, \forall i \in [0,r-1],\\
    \label{integer2} K_i \in & \mathbb{Z}^+, \forall i \in [0,r-1],\\
    &{\rm Constraint \, (\ref{equation1})}.
\end{align}
\end{subequations}

For \textit{Case} 2 of APCC, one only needs to adjust constraints (\ref{inequation3}) and (\ref{inequation6}) according to the relationship between $K_i$ and $H_i$, which does not affect the subsequent methods employed. Consequently, for the sake of convenience in expression, we will focus on \textit{Case} 1 of APCC in the following parts of this section, without loss of generality.

\subsection{APCC Without Cancellation}
If the cancellation of completed sets is not considered, we first denote the delay of one worker to continuously complete $m$ subtasks as $T_m$, and derive its CDF from (\ref{DelayT}) as:
\begin{align} \label{mTaskDelay}
    \mathbb{P}[T_m \leq t] = \left\{
    \begin{array}{ll}
         1 - e^{-\mu (\frac{t}{m} - a)}, & {\rm if} \,\, t \geq ma,  \\
         0 & {\rm otherwise}.
    \end{array}  \right.
\end{align} 
Since computations on workers are independent, $\mathbb{E}[R_i(t_i)]$ can be written as:
\begin{align}\label{ExpOfXt}
    \mathbb{E}[R_i(t_i)] = \sum_{n=0}^{N-1} \mathbb{E}[\mathbb{I}_{\{T_{i+1} \leq t_i\}}]  = N \cdot \mathbb{P}[T_{i+1} \leq t_i],
\end{align}
where $\mathbb{I}_{\{x\}}$ denotes the indicator function that equals $1$ if event $x$ is true and equals $0$ otherwise. $\mathbb{P}[T_{i+1} \leq t_i]$ is given by (\ref{mTaskDelay}).

Substituting (\ref{ExpOfXt}) into $\mathcal{P}1-2$, we find (\ref{inequation6}) is covered by (\ref{inequation3}) and obtain the following optimization problem:
\begin{subequations}
\begin{align}
    \mathcal{P}2-1: & \mathop{\min}_{\{\boldsymbol{K}, z\}} \, z\\
    \label{inequation5} \mathbf{s.t.} \, & d (K_i+L-1)+1 - N[1-e^{-\mu(\frac{t_i}{i+1}-a)}] \leq 0,\notag\\
    &\forall i \in [0,r-1],\\
    &{\rm Constraints \, (\ref{equation1}), (\ref{inequation4}), (\ref{integer2})}.
\end{align}
\end{subequations}
As $\mathcal{P}2-1$ shows, it is a mixed integer non-linear programming (MINLP) problem, which is usually NP-hard. Although its optimal solution can be found by the Branch and Bound (B\&B) algorithm \cite{lawler1966branch}, the computational complexity is up to $O\left((\frac{N}{d})^r\right)$, which means the B\&B algorithm becomes extremely time-consuming when either $N$ or $r$ are large.

\begin{algorithm}[t]

\caption{MVD}\label{MVD}
\KwIn{An arbitrary feasible solution: $\boldsymbol{K} = \{K_i, i\in [0,r-1]\}$}

\KwOut{The optimal solution: $\boldsymbol{K^*} = \{K_i^*, i\in [0,r-1]\}$}

\Do{$z_{MVD} < z^*$}{
Substitute $\boldsymbol{K}$ into the original problem and obtain the following convex optimization problem.
\begin{subequations}
\begin{align}
    \mathcal{P}:  \mathop{\min}_{\{\boldsymbol{t},z\}} \, z
    \quad \mathbf{s.t.} \, {\rm (\ref{inequation4}),(\ref{inequation5})}. \notag
\end{align}
\end{subequations}

Obtain the solution $\{\boldsymbol{t}^*,z^*\}$ to $\mathcal{P}$ by solving the Karush-Kuhn-Tucker (KKT) conditions.

Derive $z^* = \mathop{\max}_{i\in[0:r-1]} t_i^*$ and assume $t_j^* = z^*, j\in[0:r-1]$ without loss of generality.

Initialization: $\boldsymbol{K}_{MVD} = \boldsymbol{K}$ and $z_{MVD} = z^*$.

\For{$l = [0:r-1],l \neq j$}{
$\boldsymbol{K}_{temp} = \boldsymbol{K}$, $K_{j,temp} = K_j - 1$, 
$K_{l,temp} = K_l + 1$.

Substitute $\boldsymbol{K}_{temp}$ into the original problem and obtain the corresponding $z^*_{temp}$ like Step 2-3.

\If{$z^*_{temp} <  z_{MVD}$}{
$\boldsymbol{K}_{MVD} = \boldsymbol{K}_{temp}$, $z_{MVD} = z^*_{temp}$.
}
}

$\boldsymbol{K} = \boldsymbol{K}_{MVD}$.

}

\KwResult{$\boldsymbol{K}^* = \boldsymbol{K}$ is the optimal solution.}

\end{algorithm}

Accordingly, to efficiently obtain an optimal solution, we propose the maximum value descent (MVD) algorithm shown in Algorithm \ref{MVD}. The key idea of the MVD algorithm is to iteratively update the input solution $\boldsymbol{K} = \{K_i, i\in [0:r-1]\}$ by adjusting $K_i$ for the set that corresponds to the maximum value descent of the objective function $z$. In the MVD algorithm, each do-while loop can be regarded as one update, and $K_j$ in Step 7 constantly approaches the optimal $K_j^*$. Once reduced in an update, $K_j$ will not increase because the objective function $z$ must decrease in each update. When the updating process terminates, the optimal solution $\boldsymbol{K}^*$ is exactly the obtained $\boldsymbol{K}$ in the last update. Furthermore, the MVD algorithm has a computational complexity of $O\left(\frac{Nr}{d}\right)$, as the number of do-while loops is determined by constraint (\ref{inequation6}).


Besides, the MVD algorithm can be executed quickly by selecting a sufficiently good partitioning solution as input. It should be noted that after relaxation and cancellation of the integer constraint in (\ref{integer2}), $\mathcal{P}2-1$ can be transformed into a convex problem as follows:
\begin{subequations}
\begin{align}
    \mathcal{P}2-2: & \mathop{\min}_{\{\boldsymbol{K},z\}} \, z\\
    \mathbf{s.t.} \, & {\rm Constraints \, (\ref{inequation5}), (\ref{equation1}), (\ref{inequation4})},\\
    &K_i > 0, \forall i \in [0,r-1].
\end{align}
\end{subequations}
and the optimal solution is given in Proposition \ref{OptimalK_1}.  

\begin{proposition}\label{OptimalK_1}
For given $(N, K, L, d, r, \mu, a)$, the closed-form expressions of the optimal solution $\boldsymbol{K}^{[Prop1]}$ and corresponding delay $\boldsymbol{t}^{[Prop1]}$ to $\mathcal{P}2-2$ are
\begin{subequations}
\begin{align}
    &\sum_{i=0}^{r-1} e^{- \mu (\frac{z^*}{i+1} - a)} = r -\frac{d(K+rL-r)+r}{N},\\
    &t^{[Prop1]}_{i} = z^*, K^{[Prop1]}_{i} = \frac{N}{d}[1 - e^{- \mu (\frac{z^*}{i+1} - a)}] -\frac{1}{d}-L+1.\notag
\end{align}
\end{subequations}
\end{proposition}
\textit{Proof}. See Appendix B.

Due to the convexity of $\mathcal{P}2-2$, the Euclidean distance between $\boldsymbol{K}^{[Prop1]}$ and the optimal solution $\boldsymbol{K}^*$ of $\mathcal{P}2-1$ is small. Therefore, it is recommended to use a rounded result of $\boldsymbol{K}^{[Prop1]}$ as the input for the MVD algorithm.


\subsection{APCC With Cancellation}
If the cancellation of completed sets is considered, a worker may be cancelled in a certain set but successfully returns results in time for the subsequent sets. For example, worker $n$ may be a straggler for set $i$ but completes its assigned subtask and returns the result in time for the next set $(i+1)$ due to the cancellation. Such situations make it quite difficult to derive and analyze the expectation of $R_i(t)$ as in the previous Section V.B, because the impact of the cancellation of the previous set on the delay of non-straggling workers in subsequent sets need to be considered. Therefore, we provide the following alternative perspective to simplify this problem.

Note that if set $i$ is the last completed one, the entire task is completed when the last needed result for this set is received. Thus, we define the delay of set $i$ as $T_{i}^{[e]}$ and aim to minimize $\mathop{\max}_{i\in[0:r-1]} \mathbb{E}[T_{i}^{[e]}]$. To derive $\mathbb{E}[T_{i}^{[e]}]$, consider that there are still $N-H_i+1 = N-d(K_i+L-1)$ workers computing the last result for set $i$ when other sets are finished. Once any one of these workers returns the first result, this set and the entire task will be completed. Accordingly, the CDF of $T_{i}^{[e]}$ can be written as follows:
\begin{align}
    \mathbb{P}\left[T_{i}^{[e]}\leq t\right]  = 1 - \left(1- \mathbb{P}\left[T_{i+1} \leq t\right]\right)^{N-d(K_i+L-1)}\notag\\
    = \left\{
    \begin{array}{ll}
         1 - e^{-\mu (N-d(K_i+L-1)) (\frac{t}{i+1} - a)}, & {\rm if}\, t \geq (i+1)a,  \\
         0 & {\rm otherwise},
    \end{array}  \right.
\end{align}
where $T_{i+1}$ is the delay needed to complete $(i+1)$ subtasks for one worker, shown previously in (\ref{mTaskDelay}). 
Then we have 
\begin{align}
    \mathbb{E}[T_{i}^{[e]}] = \frac{i+1}{\mu[N-d(K_i+L-1)]} + a(i+1).
\end{align}
By further adding an extra optimization variable $z$ to substitute $\mathop{\max}_{i\in[0:r-1]} \mathbb{E}[T_{i}^{[e]}]$, the optimization problem can be formulated as:
\begin{subequations}
\begin{align}
    \mathcal{P}3-1: \mathop{\min}_{\{\boldsymbol{K}, z\}} \, &z \\
    \mathbf{s.t.} \, \label{inequation7} &\frac{i+1}{\mu\left[N-d(K_i+L-1)\right]} + a(i+1) -z \leq 0, \notag\\
    &\forall i \in [0:r-1],\\
    & {\rm Constraints \, (\ref{equation1}), (\ref{inequation6}), (\ref{integer2})}. 
\end{align}
\end{subequations}

Note that $\mathcal{P}3-1$ is a MINLP problem similar to $\mathcal{P}2-1$ and has an $O\left((\frac{N}{d})^r\right)$ computational complexity to solve if using B\&B algorithm. However, after relaxation and cancelling the integer constraint in (\ref{integer2}), $\mathcal{P}3-1$ can also be transformed into a convex problem as:
\begin{subequations}
\begin{align}
    \mathcal{P}3-2: & \mathop{\min}_{\{\boldsymbol{K}, z\}} \, z\\
    \mathbf{s.t.} \, & {\rm Constraints \, (\ref{inequation7}), (\ref{equation1}), (\ref{inequation6})},\\
    &K_i > 0, \forall i \in [0,r-1],
\end{align}
\end{subequations}
and optimal solution is given in Proposition \ref{OptimalK_2}.

\begin{proposition}\label{OptimalK_2}
For given $(N, K, L, d, r, \mu, a)$, the closed-form expression of the optimal solution $\boldsymbol{K}^{[Prop2]}$ to $\mathcal{P}3-2$ is
\begin{subequations}
\begin{align}
    & \sum_{i=0}^{r-1} \frac{i+1}{z^*-a(i+1)} = \mu \left[rN- d(K+rL-r)\right],\\
    &K_{i}^{[Prop2]} = \frac{N}{d} - \frac{i+1}{d\mu[z^* - a(i+1)]}-L+1.
\end{align}
\end{subequations}
\end{proposition}
\textit{Proof}. See Appendix C.

Consequently, MVD algorithm is used again to solve $\mathcal{P}3-1$ with a computational complexity of $O(\frac{Nr}{d})$, and the rounded result of $\boldsymbol{K}^{[Prop2]}$ is recommended to be used as the input. 

\section{Simulation Results}
In this section, we leverage simulation results to evaluate the performance of APCC in terms of task completion delay and compare it with other state-of-the-art coded computing strategies, including LCC\cite{LCC}, LCC with multi-message communications (LCC-MMC)\cite{ozfatura2020straggler}, and BACC\cite{BACC}. Additionally, we analyze the impact of the number of partitioned sets $r$ and the number of colluding workers $L$ on the task completion delay of APCC. 

In simulations, the entire task is given, leading to a constant computation load for the entire task. In this scenario, we aim to compare the entire task completion delay across various task division and coded computing strategies, illustrating the delay performance improvements introduced by APCC.
We assume that the computation delay $T_0$ of a single worker to complete the entire task follows a shifted exponential distribution, which is modeled as:
\begin{align}
    \mathbb{P}[T_0 \leq t] = \left\{
    \begin{array}{ll}
         1 - e^{-\mu_0 (t - a_0)}, &{\rm if} \, t \geq a_0,  \\
         0 & {\rm otherwise},
    \end{array}  \right.
\end{align}
then the computation delay $T$ of a single worker to complete one subtask follows:
\begin{align}\label{DelayT_2}
    \mathbb{P}[T \leq t] = \left\{
    \begin{array}{ll}
         1 - e^{-\mu_0 (Kt - a_0)}, & {\rm if} \,t \geq \frac{a_0}{K},  \\
         0 & {\rm otherwise},
    \end{array}  \right.
\end{align} 
where $K$ denotes the task division number, which may vary depending on the chosen coded computing strategies.
The parameter $a_0$ is set to $0.5$ seconds, and $\mu_0$ is set as $\frac{1}{10 a_0}$. In APCC, $\{K_i\}_{i=0}^{r-1}$ correspond to the number of subtasks in each set before encoding, and their value are obtained by MVD algorithm. Then, $5\times 10^4$ Monte Carlo realizations are run to obtain the average completion delay of the entire task, and the simulation codes are shared here\footnote{code link: https://github.com/Zemiser/APCC}.
Note that by comparing (\ref{DelayT_2}) with (\ref{DelayT}), we have $\mu = K \mu_0$ and $a = \frac{a_0}{K}$, and can further derive the distribution of $T_m$ in (\ref{mTaskDelay}).

The benchmarks involved in this section are as follows:
\subsubsection{APCC}
APCC is our proposed coded computing strategy in this paper. It first divides the entire task into $K$ subtasks and then partitions them into $r$ sets with different sizes. The number of subtasks in set $i, i\in[0,r-1]$ is denoted as $K_i$, which satisfies $\sum_{i=0}^{r-1}K_i = K$. After that, each set is encoded into $N$ subtasks assigned to the $N$ workers. Consequently, each worker is assigned $r$ subtasks. For Case 1 of APCC, the set $i$ is recovered when the master has received $d(K_i +L-1)+1$ results, and the entire task is completed when all sets are recovered.
\subsubsection{LCC}
LCC proposed in \cite{LCC} divides the entire task into $K'$ subtasks and then encodes them into $N$ subtasks assigned to $N$ workers. Each worker in LCC is assigned one subtasks. Therefore, the entire task is completed when the master has received $d(K'+L-1)+1$ results. $L=0$ means the absence of a requirement for privacy preservation. We assume that the number of workers $N$ is greater than $dK'-1$ to facilitate our analysis. Consequently, when $L=0$, the recovery threshold is defined as $d(K'-1)+1$ instead of $N - \lfloor N/K' \rfloor + 1$ according to \cite{LCC}.
\subsubsection{LCC-MMC} MMC proposed in \cite{ozfatura2020straggler} is an another approach to utilize the computing results of straggling workers except for the hierarchical structure. It also achieves partial returning of results from workers through a more granular task division.  Specifically, LCC-MMC divides the entire task into $K^{LM}$ subtasks and then encodes them into $Nr$ subtasks. Each worker in LCC-MMC is assigned $r$ subtasks and the entire task is completed when the master has received $d(K^{LM}-1)+1$ results. However, LCC-MMC can not preserve the privacy of input data because multiple encoded data from the same encoding function is sent to a worker, which is different from the case of APCC where $r$ subtasks assigned to the same worker are generated by $r$ different encoding functions $\{g_i(x)\}_{i=0}^{r-1}$.
\subsubsection{BACC}
The BACC strategy, as introduced in \cite{BACC}, offers approximate results with improved precision achievable by increasing the number of return results from workers. It shares a task division structure identical to LCC, partitioning the task into $K'$ subtasks and then further encoding them into $N$ subtasks. Each worker in BACC is assigned one such subtask. 

To ensure fairness, and all strategies employ an identical number of workers and distribute an equivalent computation loads for a single worker. Assuming that the computation loads of the entire task are $O(\gamma)$, then each subtask $f(D_k)$ in APCC has a computation load of $O(\frac{\gamma}{K})$, and the computation loads of each worker in APCC are $O(\frac{\gamma r}{K})$ because there are $r$ partitioned sets. Similarly, we can derive that the computation loads of each worker in LCC, BACC and LCC-MMC are $O(\frac{\gamma}{K'})$, $O(\frac{\gamma}{K'})$ and $O(\frac{\gamma r}{K^{LM}})$, respectively. In order to ensure that each worker in these schemes performs an identical fraction of the entire task as APCC, we have 
\begin{align}
K' =  K^{LM}/r = K/r .
\end{align}

Due to the different applicability of various coded computing strategies, we will first conduct a comprehensive analysis and comparison of APCC alongside other strategies within the following three scenarios: 1) Accurate results with $L$ colluding workers ($L>0$); 2) Accurate results without colluding workers ($L=0$); 3) Approximated results. Finally, we study the impact of the parameters $r$ and $L$ on the delay performance of APCC.

\subsection{Accurate Results With $L$ Colluding Workers ($L>0$)}
In this scenario, we consider the following three benchmarks: LCC, APCC without cancellation and APCC with cancellation. For fair comparison, the computation load of workers should be set the same, so we have $K' = K/r$.

\begin{figure}
    \centering
    
    \includegraphics[scale = 0.44]{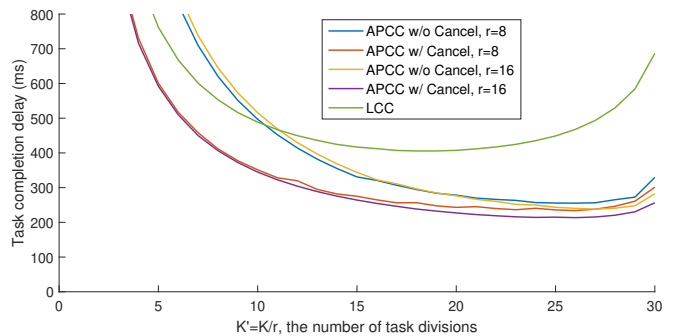
    }
  
    \caption{Delay performance comparison between APCC and LCC for accurate results with $L$ colluding workers ($L>0$). Settings: $N = 200, L = 20, d = 4$. The partitioning strategy $\{K_i\}$ of APCC is obtained by the proposed MVD algorithm. $r$ is the number of partitioned sets.}
    \label{Accurate_L_exceed_0_p1}
\end{figure}

As shown in Fig. \ref{Accurate_L_exceed_0_p1}, the average completion delay of the entire task $\{f(D_k)\}_{k=0}^{K-1}$ first decreases and then increases with the task division number $K$, indicating the existence of an optimal division that minimizes the delay. This trade-off arises from balancing the computation load of each worker and the minimum number of workers needed to recover $\{f(D_k)\}_{k=0}^{K-1}$. On the one hand, as the division number decreases, the computation load of each subtask increases, which leads to longer computation delays for each worker due to the increased workload. Although the number of workers waiting for results decreases, the increase in load negates this advantage. On the other hand, while the division number approaches the maximum, 
as illustrated in the inequality (\ref{K Constr}), 
the number of workers that the master needs to wait for approaches $N$, making the straggling effect a bottleneck for performance and increasing the delay. The zigzag fluctuations in the curve are mainly due to the integer values of the partitioning numbers.

Note that the primary metric for evaluating different schemes in our study is the minimum task completion delay under different division numbers, as depicted in Fig. \ref{Accurate_L_exceed_0_p1}. This is because the division number $K' = \frac{K}{r}$ corresponds to the division of computation function inputs, which is typically a high-dimensional matrix. As such, $K'$ can be adjusted flexibly in most cases. Therefore, the minimum achieved task completion delay is the main focus of our analysis.

\begin{figure}
    \centering
    \includegraphics[scale = 0.44]{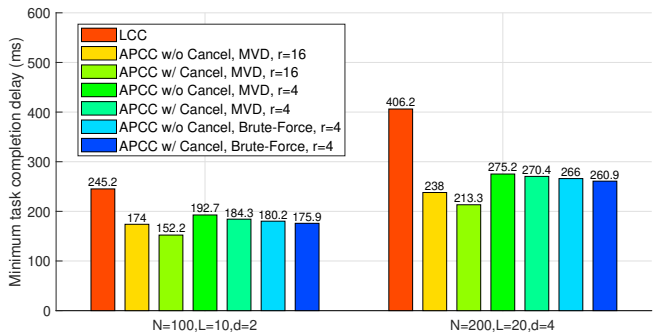}
    \caption{APCC v.s. LCC. Minimum task completion delay achieved by all possible task divisions $K' = K/r$, applied to accurate results with $L$ colluding workers ($L>0$).}
    \label{APCCvsLCC}
\end{figure}


Fig. \ref{APCCvsLCC} compares APCC and LCC in terms of the minimum task completion delay. In these benchmarks, 'Brute-Force' refers to a partitioning strategy derived from an exhaustive search across all possible values of $\{K_i\}$. Due to the highly complex traversal search, the brute-force results are only provided for scenarios with a smaller number of sets ($r=4$). 
Fig. \ref{APCCvsLCC} illustrates that both APCC with and without cancellation yield sufficient reductions in task completion delay compared to LCC. For instance, when $N=100, L=10, d=2, r=16$, and the partitioning strategy obtained from MVD algorithm is utilized, APCC with and without cancellation achieve delay reductions of $41.4\%$ and $47.5\%$, respectively, compared to LCC. Moreover, the comparison with the 'Brute-Force' benchmarks show that the partitioning strategy $\{K_i\}$ obtained through the MVD algorithm is near-optimal.


\subsection{Accurate Results Without Colluding Workers ($L=0$)}
In this scenario, we evaluate four benchmarks: LCC, LCC-MMC, APCC with and without cancellation. Among these, only LCC does not consider partial results from straggling workers. Similar to Subsection IV.A, we set $K' = K^{LM} = K/r$, with $K^{LM}$ representing the task division number for LCC-MMC.

\begin{figure}
    \centering
    \includegraphics[scale = 0.44]{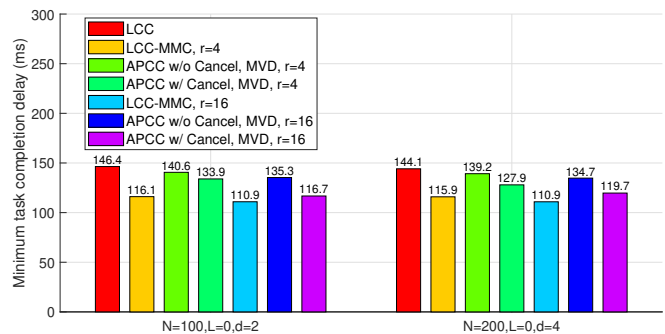}
    \caption{APCC v.s. LCC and LCC-MMC. Minimum task completion delay achieved by all possible task divisions $K' =K^{LM}/r= K/r$, applied to accurate results without colluding workers ($L=0$).}
    \label{APCCvsLCCandLCCMMC}
\end{figure}

In Fig. \ref{APCCvsLCCandLCCMMC}, both LCC-MMC and APCC effectively reduce task completion delay compared to LCC. Specifically, when $r$ is large enough, APCC with cancellation closely approaches the performance of LCC-MMC. This similarity arises because in both APCC and LCC-MMC, the master utilizes nearly all computing results from workers when divided subtasks are sufficiently small. Fig. \ref{APCCvsLCCandLCCMMC} also illustrates that when privacy is not a concern, MMC is a viable method to reduce the delay in coded computing.

Comparing to Fig. \ref{APCCvsLCC}, we observe that the absence of colluding workers limits the potential for delay optimization. For instance, with parameters $N = 100$, $L = 0$, $d = 2$, and $r = 16$, APCC with cancellation achieves only a $20.3\%$ delay reduction compared to LCC.

\subsection{Approximated Results}
In this subsection, we compare the task completion delay of BACC and the case 2 of APCC, which both can provide approximated results with fewer workers than the recovery threshold. To ensure uniform worker computation load, we also set $K' = K/r$, as in our previous analysis. Furthermore, since BACC shares an identical task division structure with LCC, we employ a smaller recovery threshold of the same form as LCC to evaluate its delay performance. For instance, when the recovery threshold $d(K' + L - 1) + 1$ exceeds $N$, a reduced uniform recovery threshold $\frac{d}{2}(K' + L - 1) + 1$ below $N$ can be employed for both BACC and APCC. 

\begin{figure}
    \centering
    \includegraphics[scale = 0.44]{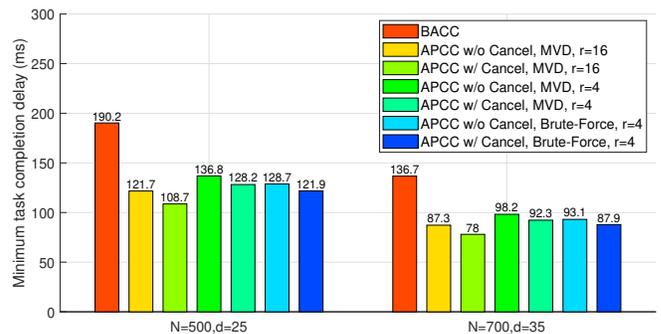}
    \caption{APCC v.s. BACC. Minimum task completion delay achieved by all possible task divisions $K'= K/r$, applied to approximated results.}
    \label{APCCvsBACC}
\end{figure}

As shown in Fig. \ref{APCCvsBACC}, the hierarchical task partitioning and the cancellation of completed sets in APCC yield sufficient delay performance improvement. Compared to BACC, the proposed MVD algorithm for APCC achieves up to $42.9\%$ delay reduction. Note that in this scenario, both APCC and BACC can obtain approximated results with fewer returned results, while LCC for accurate computation fails to work when $K'$ is larger than $20$ in the two cases of Fig. \ref{APCCvsBACC}, as the recovery threshold of LCC needs to be larger than $d(K' + L-1)+1$.

\subsection{Impact of $r$ and $L$ on the Performance of APCC}

\begin{figure}
    \centering
    \subfigure[Delay of APCC v.s. $r$]{
    \includegraphics[scale = 0.41]{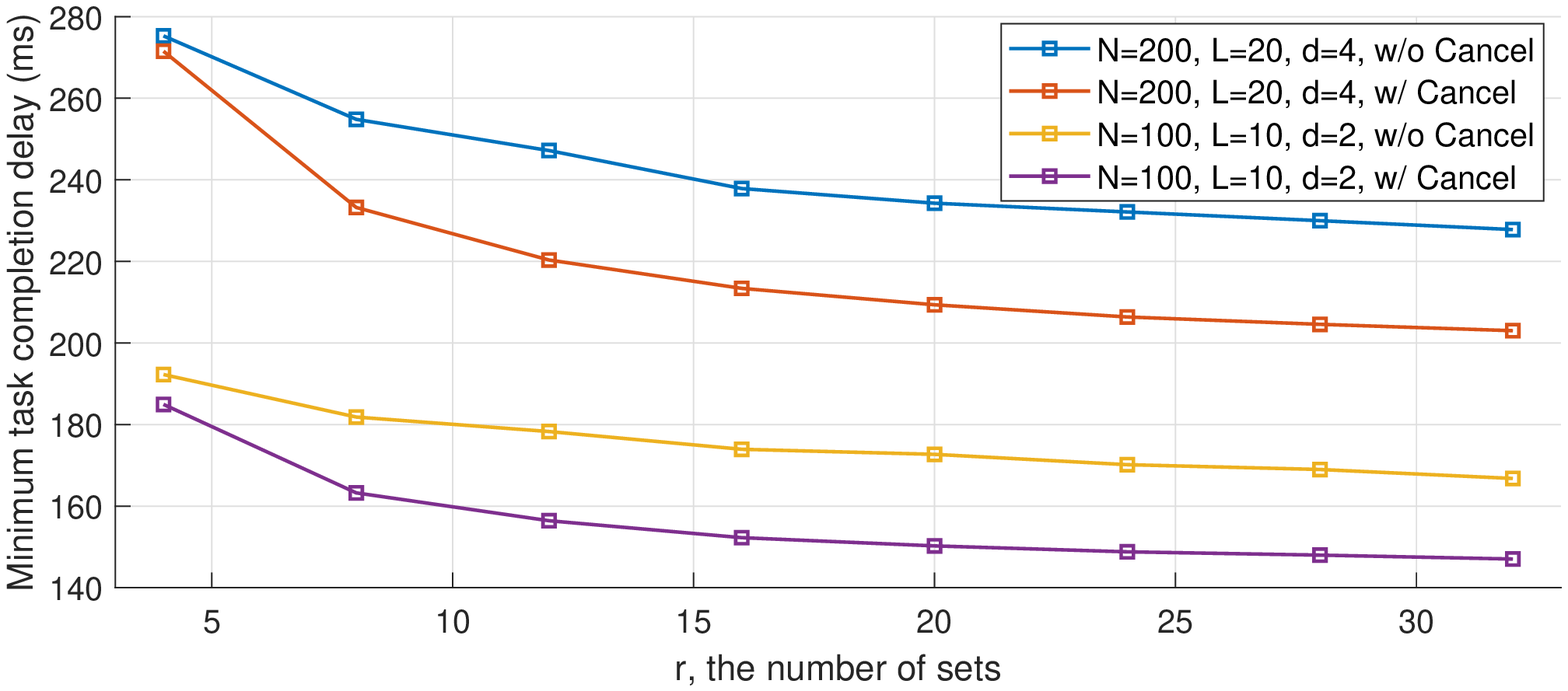}
    }
    \subfigure[$N=200, d=4,r=12$]{
    \includegraphics[scale = 0.41]{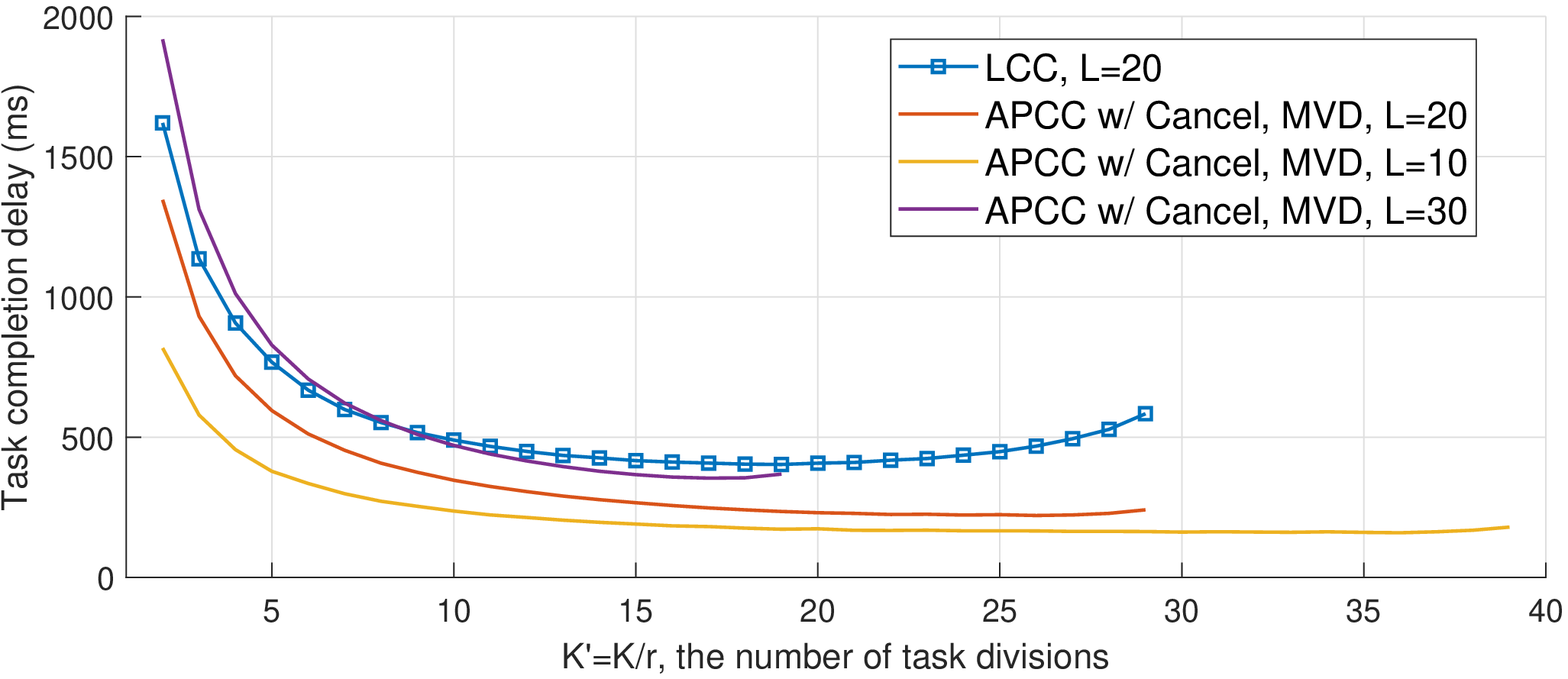}
    }
    \caption{Delay performance of APCC with different $r$ and $L$.}
    \label{Delay_vs_r_L}
\end{figure}

The impact of the hierarchical partitioning number of sets $r$ on the task completion delay of APCC is illustrated in Fig. \ref{Delay_vs_r_L}(a). It is observed that a larger number of sets $r$ results in a smaller computation delay, which is consistent with the results shown in previous figures. The reduction in delay can be attributed to the fact that a larger $r$ implies a smaller computation load for each subtask in the hierarchical structure, and the difference in computation load between fast and slow workers can be described more precisely. Consequently, the proposed MVD algorithm can better utilize the computing results of straggling workers to reduce delay.
Furthermore, Fig. \ref{Delay_vs_r_L}(a) indicates that the benefit of increasing $r$ has a boundary effect, which corresponds to the upper bound of benefit brought by the granularity refinement of task divisions.

Recall that $L$ denotes the maximum number of colluding workers that a coded computing scheme can tolerate. The value of $L$ can serve as an indirect indicator of the level of privacy preservation offered by the scheme. Specifically, a larger value of $L$ corresponds to more stringent privacy protection and a higher tolerance for colluders. It is demonstrated in Section IV.B that complete data privacy can be achieved as long as the number of colluders remains below $L$.

Fig. \ref{Delay_vs_r_L}(b) illustrates the impact of the number of colluding workers $L$ on the trade-off between delay and privacy preservation. It is worth noting that, for a fixed $K'$, increasing the value of $L$ leads to a larger recovery threshold $H$ for the original subtasks, which results in a longer task completion delay. Moreover, as demonstrated in (\ref{K Constr}), choosing a larger value of $L$ restricts the maximum number of task divisions. Consequently, the range of $K'$ values corresponding to the plotted curves in Fig. \ref{Delay_vs_r_L}(b) varies with $L$.

\section{Conclusion}
In this paper, we have investigated a distributed computing system that consists of one master and multiple workers. We have first proposed an adaptive privacy-preserving coded computing (APCC) strategy, which is suitable for diverse task scenarios and computation functions. APCC adaptively provides accurate or approximated results with controllable error according to the form of computation functions, and the computation process keeps numerically stable. We have rigorously proved the optimality of APCC in terms of encoding rate. The complete privacy preservation of input data has also been proved.

We have further provided a low-complexity maximum value descent (MVD) algorithm to optimally solve the hierarchical task partitioning problem in APCC, with and without considering cancellation, aiming at minimizing task completion delay. The cancellation is our proposed operation aiming to further accelerate computation by timely cancelling the completed tasks. Extensive simulations have demonstrated that APCC outperforms the state-of-the-art coded computing strategies by a range of $20.3\%$ to $42.9\%$ in terms of task completion delay.



\appendices
\section{Proof of the inequality (\ref{LCC_optimality})}


In this appendix, the proof for the optimal recovery threshold of LCC \cite{LCC} to guarantee decodablity is briefly introduced to enhance the clarity of the inequality (\ref{LCC_optimality}). Additionally, we provide an alternative and more concise proof for the construction of the involved multilinear function.

To simplify the proof, a weakened result under the condition of multilinearity is first derived. After that, in order to extend to the case of a general polynomial function, a construction of multilinear functions based on polynomial functions is provided. 
The definition of the multilinear function is as follows:
\begin{definition}
For a multilinear function $f(D_1,D_2,\dots D_d)$ defined on $\mathbb{V}$ with degree $d$, $D_1,D_2,\dots D_d$ are its $d$ input variables, and $f$ is linear with respect to each variable.
\end{definition}


Under the assumption of the multilinearity of $f$, the optimal recovery threshold is provided in Lemma 1 of \cite{LCC} as:
\begin{lemma} \label{Multilinear lemma}
{\rm\cite{LCC}} Consider an $(N,S,L,f)$ coded computing problem, where $N$ is the number of workers, $S, L$ is the maximum number of stragglers and colluding workers that can be tolerated, respectively. $f$ is a multilinear function, the degree of $f$ is $d$, and the number of the equally divided input data is $K$. The optimal recovery threshold for linear coded computing schemes, denoted by $H^*$, is defined as:
\begin{equation}
    H^* \triangleq \left\{
    \begin{array}{ll}
         d(K+ L -1)+1, & {\rm if} \,\, L>0,  \\
         \min\{d(K-1)+1, \, N - \lfloor N/K \rfloor + 1\} & {\rm if} \,\,   L=0.
    \end{array}  \right.
\end{equation}
\end{lemma}
In order to generalize to the case of polynomial functions, a construction method of multilinear functions is given in Lemma 4 of \cite{LCC} as follows:
\setcounter{lemma}{3}
\begin{lemma}\label{Construction of Multilinear}
{\rm \cite{LCC}} For a general polynomial function $f$ with degree $d$, $f'$ is a $d$-variable multilinear polynomial function constructed based on $f$ and satisfies:
\begin{align}
    f'(D_1,D_2,\dots,D_d) = \sum_{\mathcal{T} \subseteq[1:d]} [ (-1)^{|\mathcal{T}|} f(\sum_{k\in \mathcal{T}} D_k) ],
\end{align}
where $f'$ is linear with respect to each input variable, $\mathcal{T}$ is a subset of the set $[1:d]$ and the degree of $f'$ also equals $d$.
\end{lemma}

Though Lemma \ref{Construction of Multilinear} has been proved in \cite{LCC}, here we provide an alternative and more concise proof.
In order to prove Lemma \ref{Construction of Multilinear}, we need to prove the order of each variable in $f'$ is at most 1 due to the multilinearity of $f'$. Therefore, if the coefficients of higher-order terms in the multilinear polynomial function $f'$ equal 0, the proof is completed.

For any $j\in[1:d]$, we use $h(D_j)$ to denote a general higher-order term in $f'$. In $h(D_j)$, the order of $D_j$ is greater than 1, and $h(D_j)$ consists of $\{D_j,D_{j_1},D_{j_2},\dots,D_{j_m}\}$ through multiplication.
Also, the number of subsets $\mathcal{T}$ that have $\{j,j_1,j_2,\dots,j_m\}$ is $2^{(d-m-1)}$, and the constant coefficients of $h(D_j)$ are the same for these different $\mathcal{T}$ in the calculation result of $f(\sum_{k\in \mathcal{T}} D_k)$. We only need to consider the impact of $(-1)^{|\mathcal{T}|}$.

Note that for $i\in[0:d-m-1]$, there are $\left(\begin{array}{c} i \\ d-m-1 \end{array} \right)$ subsets $\mathcal{T}$ that meet above conditions and include extra $i$ variables except for $\{j,j_1,j_2,\dots,j_m\}$. Consequently, any coefficient of $h(D_j)$, denoted by ${\rm Coeff}_j$, can be obtained as:
\begin{align}
    &{\rm Coeff}_j = \sum_{\{j,j_1,j_2,\dots,j_m\} \subseteq \mathcal{T},\mathcal{T} \subseteq[1:d]} (-1)^{|\mathcal{T}|}, \nonumber \\
    &= \sum_{i=0}^{d-m-1} \left(\begin{array}{c} d-m-1 \\ i \end{array} \right) (-1)^{m+1+i}, \nonumber \\
    &= (-1)^{m+1} \cdot \sum_{i=0}^{d-m-1}\left[ \left(\begin{array}{c} d-m-1 \\ i \end{array} \right) 1^{d-m-1-i} (-1)^i \right], \nonumber \\
    &= (-1)^{m+1} \cdot (1-1)^{d-m-1} = 0,
\end{align}
which completes the proof of Lemma \ref{Construction of Multilinear}.

Based on above two lemmas, Lemma \ref{Multilinear lemma} can be extended to the case of general polynomial \cite{LCC}. Moreover, the actual number of results returned by workers equals $(N-S)$, which must be larger than the recovery threshold. Consequently, to guarantee the decodability for general polynomial coded computing, $N-S \geq H^*$ should hold, and thus the formula (\ref{LCC_optimality}) is derived.






\vspace{-5pt}
\section{Proof of Proposition \ref{OptimalK_1}}
The Lagrangian of $\mathcal{P}2-2$ is given by:
\begin{align}
    &\mathcal{L}(\boldsymbol{K}, \boldsymbol{t}, z, \lambda, \{\alpha_i\}, \{\beta_i\}) \notag \\
    = & z + \lambda \left(\sum_{i=0}^{r-1}K_i - K\right) + \sum_{i=0}^{r-1} \alpha_i(t_i - z)  \notag\\
    +& \sum_{i=0}^{r-1} \beta_i \left[d(K_i+L-1)+1 - N(1 - e^{-\mu(\frac{t_i}{i+1} - a)})\right],
\end{align}
where $\lambda,\{\alpha_i\}$ and $\{\beta_i\}$ are the Lagrange multipliers associated with (\ref{equation1}), (\ref{inequation4}) and (\ref{inequation5}), respectively.

The partial derivatives of $\mathcal{L}(\boldsymbol{K}, \boldsymbol{t}, z, \lambda, \{\alpha_i\}, \{\beta_i\})$ can be derived as:
\begin{subequations}
\begin{align}
    \frac{\partial \mathcal{L}}{\partial K_i} &= \lambda + d\beta_i, \quad \frac{\partial \mathcal{L}}{\partial z} = 1 -\sum_{i=0}^{r-1}\alpha_i,\\
    \frac{\partial \mathcal{L}}{\partial t_i} &= \alpha_i - \frac{\beta_i N \mu}{i+1} e^{-\mu(\frac{t_i}{i+1} - a)}.
\end{align}
\end{subequations}

The Karush-Kuhn-Tucker (KKT) conditions are written as:
\begin{subequations}
\begin{align}
    &\frac{\partial \mathcal{L}}{\partial H_i^*}=0, \frac{\partial \mathcal{L}}{\partial t_i^*}=0, \frac{\partial \mathcal{L}}{\partial z^*} = 0, \\
    &\beta_i^* \left[ d(K_i^*+L-1)+1 - N(1 - e^{-\mu(\frac{t_i^*}{i+1} - a)}) \right] = 0, \\
    &\alpha_i^*(t_i^* - z^*) = 0, \\
    &\alpha_i^*, \beta_i^* \geq 0, K_i^* > 0, \forall i \in [0:r-1].
\end{align}
\end{subequations}
by solving the KKT conditions, the optimal solution to $\mathcal{P}2-2$ is obtained, as shown in Proposition \ref{OptimalK_1}.

\section{Proof of Proposition \ref{OptimalK_2}}
Similarly, as the constraints of $\mathcal{P}3-2$ are convex, problem $\mathcal{P}3-2$ is a convex optimization problem. The Lagrangian of $\mathcal{P}3-2$ is given by:
\begin{align}
    &\mathcal{L}(\boldsymbol{K}, z, \lambda, \{\alpha_i\})
    =  z + \lambda \left(\sum_{i=0}^{r-1}K_i - K \right) \notag\\
    +& \sum_{i=0}^{r-1} \alpha_i \left[ \frac{i+1}{\mu(N-d(K_i+L-1))} + a(i+1) - z \right].
\end{align}

For the convex problems, the optimal solution $\{\boldsymbol{K}^*,z^*\}$ must satisfy the KKT conditions. By solving
\begin{subequations}
\begin{align}
    &\frac{\partial \mathcal{L}}{\partial K_i^*} = \lambda^* + \frac{\alpha_i^* d(i+1)}{\mu(N-d(K_i^*+L-1))^2} = 0,\\
    &\frac{\partial \mathcal{L}}{\partial z^*} = 1 -\sum_{i=0}^{r-1}\alpha_i^* = 0, \quad \alpha_i^* \geq 0, \\
    &\alpha_i^* \left[ \frac{i+1}{\mu(N-d(K_i^*+L-1))} + a(i+1) - z^* \right] = 0,\\
    &(N-1)/d-L+1 \geq K_i^* >0, \forall i \in [0:r-1]
\end{align}
\end{subequations}
the optimal solution to $\mathcal{P}3-2$ is obtained.


{
\bibliographystyle{IEEEtran}
\footnotesize
\bibliography{_bare_jrnl_bib}
}

\end{document}